\documentclass[sigconf]{acmart}
\usepackage{tcolorbox}

\AtBeginDocument{%
  \providecommand\BibTeX{{%
    \normalfont B\kern-0.5em{\scshape i\kern-0.25em b}\kern-0.8em\TeX}}}

\setcopyright{acmcopyright}
\copyrightyear{2021}
\acmYear{2021}
\acmDOI{10.1145/1122445.1122456}

\copyrightyear{2021} 
\acmYear{2021} 
\setcopyright{acmlicensed}\acmConference[CHI '21]{CHI Conference on Human Factors in Computing Systems}{May 8--13, 2021}{Yokohama, Japan}
\acmBooktitle{CHI Conference on Human Factors in Computing Systems (CHI '21), May 8--13, 2021, Yokohama, Japan}
\acmPrice{15.00}
\acmDOI{10.1145/3411764.3445184}
\acmISBN{978-1-4503-8096-6/21/05}

\begin{document}

\title{A Probabilistic Interpretation of Motion Correlation Selection Techniques}

\author{Eduardo Velloso}
\email{eduardo.velloso@unimelb.edu.au}
\orcid{0000-0003-4414-2249}
\affiliation{%
  \institution{The University of Melbourne}
  \streetaddress{Parkville}
  \state{Victoria}
  \country{Australia}
  \postcode{3010}
}

\author{Carlos Hitoshi Morimoto}
\affiliation{%
  \institution{University of S\~{a}o Paulo}
  \streetaddress{Butant\~a}
  \city{S\~ao Paulo}
  \country{Brazil}}
\email{hitoshi@ime.usp.br}

\renewcommand{\shortauthors}{Velloso and Morimoto}

\begin{abstract}
Motion correlation interfaces are those that present targets moving in different patterns, which the user can select by matching their motion. In this paper, we re-formulate the task of target selection as a probabilistic inference problem. We demonstrate that previous interaction techniques can be modelled using a Bayesian approach and that how modelling the selection task as transmission of information can help us make explicit the assumptions behind similarity measures. We propose ways of incorporating uncertainty
into the decision-making process and demonstrate how the concept of entropy can illuminate the measurement of the quality of a design. We apply these techniques in a case study and suggest guidelines for future work.
\end{abstract}



\maketitle

\section{Introduction}
Though the dominant interaction paradigm in current gestural user interfaces is still largely based on deictic and manipulative gestures, a different kind of interaction has been gathering interest in recent years---that of \textit{motion correlation}~\cite{velloso2017motioncorrelation}. The principle underlying this type of interaction relies on the interface presenting moving targets, each with a distinct motion pattern, which the user can mimic in order to signal the intention to select the desired target~\cite{velloso2017motioncorrelation}. The system then measures the similarity between the signal originating from the input device and the signals corresponding to the targets moving on the screen in order to determine which target (if any) is being followed.

The literature contains a wealth of examples that successfully employ the principle, including for gaze interaction with smart watches~\cite{esteves2015orbitsa,esteves2015orbitsb}, public displays~\cite{vidal2013pursuitsb,khamis2016textpursuits}, virtual reality~\cite{khamis2018vrpursuits}, and smart homes~\cite{velloso2016ambigaze}; for manual control in body-based games~\cite{carter2016pathsync,Cox2016} and smart TVs~\cite{clarke2016tracematch,clarke2017matchpoint,verweij2017wavetrace}; for head control of head-up displays~\cite{esteves2017smoothmoves}; for one-handed ring-based input~\cite{zhang2017fingorbits}; for menu selection with mice~\cite{williamson2004pointing,fekete2009motion}; among others.

Despite building upon the same principle, these works employ a wide variety of input devices, data processing pipelines, similarity measures, feedback mechanisms, and evaluation procedures. As a consequence, it is challenging to compare approaches and results in order to advance the state-of-the-art of motion correlation interaction. The standard analysis approach in related work is empirical, instantiating different designs and collecting performance data (e.g. precision, recall, selection times, and error rates) in user studies. The problem with this is that the design space of motion correlation interfaces is enormous---not only there are many design parameters to be manipulated, but these parameters are not necessarily independent.

In this paper, we propose an interpretation of motion correlation based on probability and information theory, inspired by the early work conducted by Williamson and Murray-Smith~\cite{williamson2004pointing,williamson2006continuous} and aligned with current trends in computational interaction~\cite{oulasvirta2018computational}. As such, we offer the following contributions: (1) We demonstrate that previous works on motion correlation can be modelled as a probabilistic reasoning problem; (2) we demonstrate a series of probabilistic techniques for better understanding combinations of interface design decisions; (3) we offer a case study of the application of these techniques in the context of a motion correlation technique, \textit{Orbits}~\cite{esteves2015orbitsa}. We argue that such account offers a holistic approach for understanding motion correlation and a consistent language for framing future work and advancing progress on the topic. 

\section{Related Work}

\subsection{Motion Correlation Interfaces}
A motion correlation interface presents the user $n$ targets, each moving according to its own characteristics, such as trajectory, velocity, phase, and direction. The user signals their intention to select a target by matching the motion of the desired target. When the system detects that the user has been following one of the targets, the selection of the corresponding target occurs. 
The main advantage of this technique is that users can signal their intention to select a target by mimicking its motion with their eyes, hand, head, or finger, without the need of an explicit pointing device or selection confirmation mechanism. Previous works have demonstrated a wealth of opportunities for motion correlation interfaces, including enabling calibration-free gaze interaction~\cite{vidal2013pursuitsb,velloso2016ambigaze}, implicitly calibrating eye trackers~\cite{pfeuffer2013pursuit,khamis2016textpursuits,gomez2018smoothi}, enabling cursorless multi-user gestural interaction~\cite{carter2016pathsync}, and enabling interaction with small displays~\cite{esteves2015orbitsa,esteves2015orbitsb}.

The central challenges for a motion correlation system are to detect \textit{whether} the user is trying to select a target, and if so, \textit{which} target they are trying to select. To accomplish this, system designers must carefully choose a set of parameters that together determine the performance of the system. These parameters include the number of targets, their motion characteristics (e.g. deterministic vs. random, path shape, velocity, acceleration, direction, etc.), the input modality and device (e.g. eye trackers, RGB cameras, depth cameras), the feedback modality and device (e.g. screens, projections, mechanical actuators), the similarity measure used to compare the projection of the user movements onto the input space with the target motions (e.g. Euclidean distance, Pearson's correlation, frequency analysis, etc.), the amount of data being compared (i.e. the number of samples within a window), the decision criteria for determining when a selection happens (e.g. the correlation threshold), among many others (see Velloso et al. for an overview of these parameters~\cite{velloso2017motioncorrelation}). 

The large number of parameters creates a rich design space, with many opportunities for building novel interfaces. The downside is that thoroughly exploring and understanding this design space becomes a challenge without a systematic process for comparing alternatives. Previous works have taken an empirical approach for such exploration---designing interfaces with different combinations of parameters and evaluating them through user studies. However, recruiting participants is expensive and time-consuming. In this work, we propose techniques for early design space exploration that enable us to analyse design decisions a priori.

\subsection{Probabilistic Interaction}


Any interaction with a system involves a level of uncertainty, including factors such as noise, user capabilities, and sensor limitations. Probabilistic interaction techniques, in contrast to conventional deterministic ones, explicitly model this uncertainty to enhance and improve the interaction. They typically draw from probability theory, information theory, control theory, machine learning, and other related areas.

For example, Rogers et al.~\cite{rogers2010fingercloud} employed a particle filtering approach  to improve finger touch estimation using low resolution capacitive sensing arrays, and Biswas et al.~\cite{Biswas:2013} used a Kalman filter algorithm to estimate the cursor position and the probability distribution of possible targets to improve the performance of pointing tasks for people with physical impairments. These approaches have in common the modelling of likelihoods of possible user states and use this to improve the interaction.

A more general framework for handling input with uncertainty was proposed by Schwarz et al.~\cite{schwarz2010framework}. The framework tracks the probabilities of alternative inputs, provides a mechanism to dispatch the input to appropriate interactors (or interface elements), and allows the interactors to provide feedback (mediation) or other appropriate actions to resolve ambiguous input cases. Another general framework based on information theory for modelling interaction is that of \textit{Bayesian Information Gain}, which has been demonstrated in applications such as pan and zoom navigation and file retrieval~\cite{liu2017bignav,liu2018bigfile}. The approach interprets every user action as clue for what the user is interested in, updating a probabilistic model of the information space. The system can then update its visualisation in order to maximise the expected information gain of the system. We apply ideas inspired from these approaches to the specific problem of selection through motion correlation.

\begin{figure}[ht]
    \begin{tabular}{cc}
    \includegraphics[width=0.47\columnwidth]{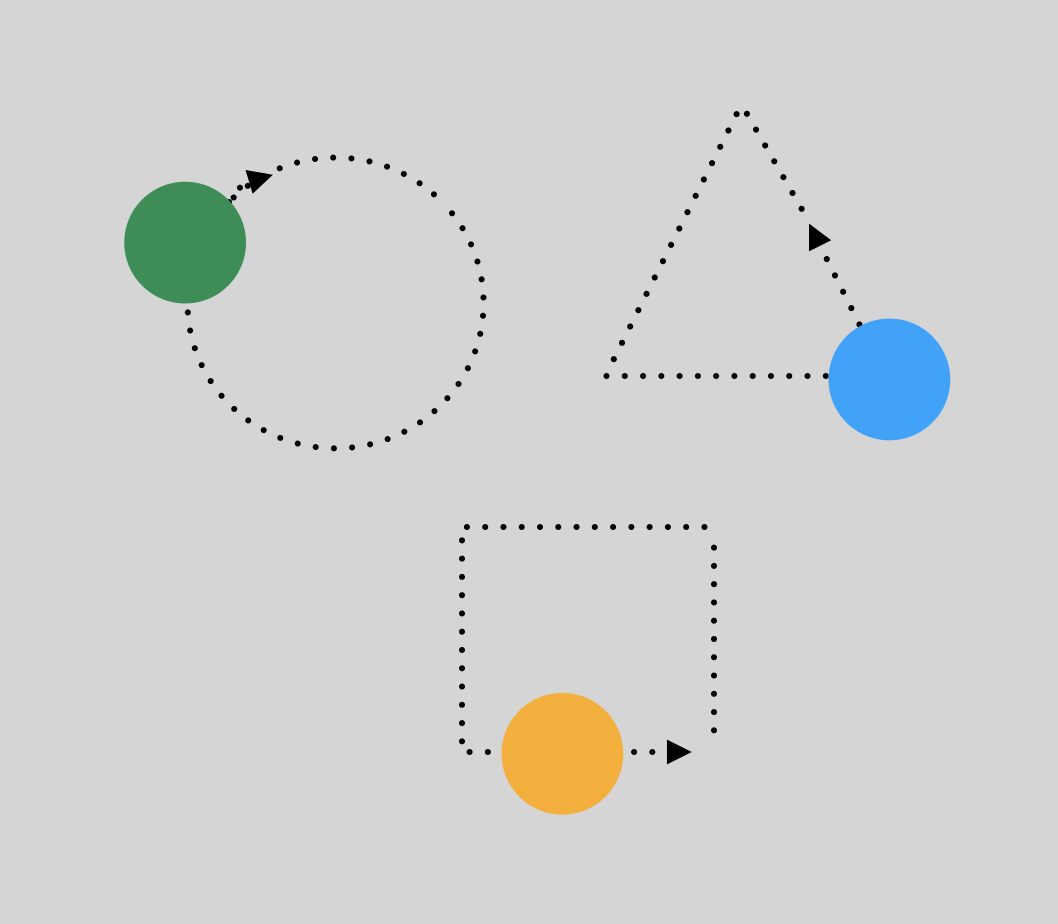} &
    \includegraphics[width=0.47\columnwidth]{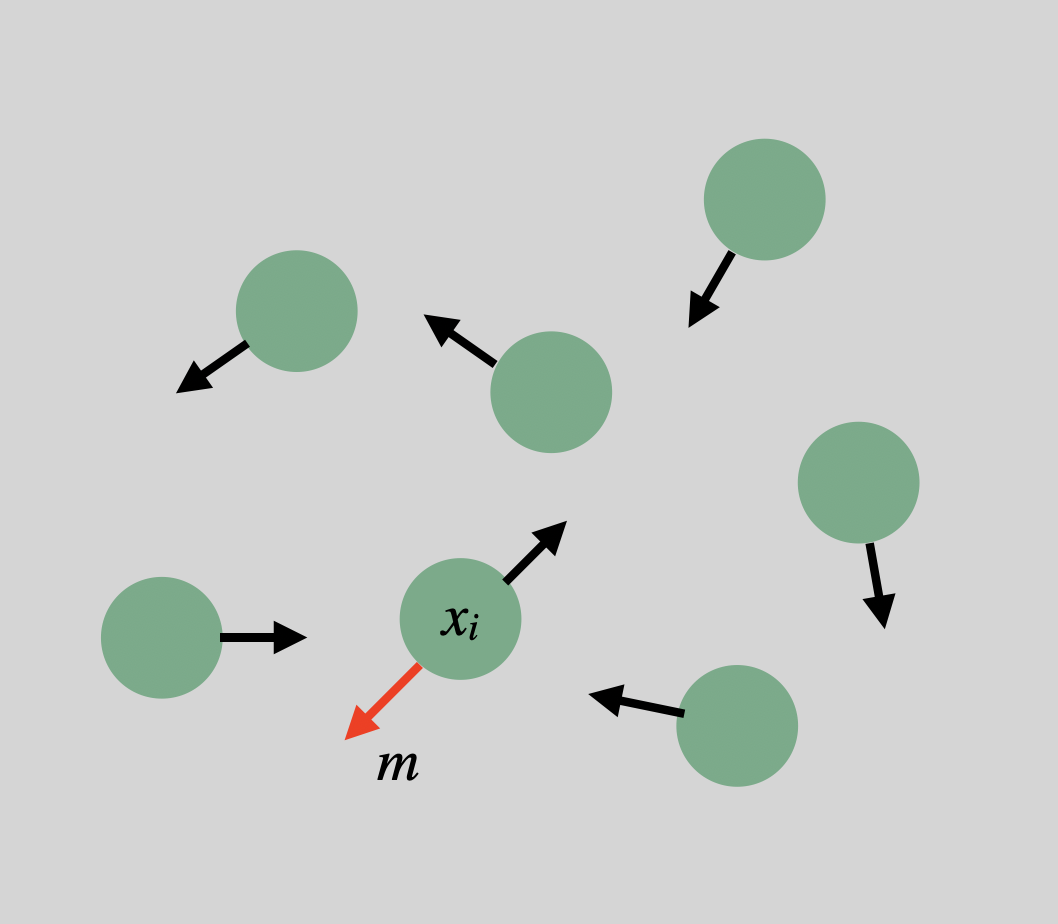} \\
    a & b
    \end{tabular}
    \caption{a) The basic motion correlation idea selects a target moving along a known trajectory using some similarity measure; and b) shows the idea of the pointing without a pointer method proposed by Williamson and Murray-Smith. Black arrows indicate the expected motion of each target. To select target $x_i$ the user must stabilise it moving the mouse in the opposite direction $m$.}
    \label{fig:Targets}
\end{figure}

In the context of motion correlation we consider the literature about probabilistic methods used in gestural user interfaces. 
A gesture can be represented as a temporal sequence of measurement samples provided by at least one position or motion sensor, for example. A particular gesture can be recognised by matching the shape trajectory described by the user's motion to expected known shapes, such as seen in Figure~\ref{fig:Targets}a. 
Sensor noise and personal motion traits are possible sources variations in the path obtained by the input device that are challenging for any gesture recognition system. 
Statistical shape analysis methods~\cite{Dryden:98} have been designed to handle such variations in the interpretation of gestures.
Active shape and appearance models are examples of such techniques used in computer vision to find objects (such as hands and faces) in images~\cite{Cootes:2004} and for recognition of handwritten characters~\cite{shi2003handwritten}.

Previous motion correlation methods have mostly taken a deterministic approach for detecting target selection. Typically, they measure the similarity between the projection of the user's movements onto the input space and each target's motion; then, if the similarity crosses a given threshold, the target is selected. 

A notable exception was the ``pointing without a pointer'' technique proposed by Williamson and Murray-Smith~\cite{williamson2004pointing}. Their idea was to present the user several targets moving according to a pseudo-random disturbance. An arrow attached to each target shows the user the direction where the target is moving to due to the disturbance. To select a target, the user must try to stabilise its motion by moving the mouse in the opposite direction of the target's arrow, as illustrated in Figure~\ref{fig:Targets}b.  The position of a target $x_i$ at time $t$ is computed by adding the mouse motion $m$ to the disturbance for $x_i$. The variance of this sum, $\sigma_{si}$, is computed over a time window $w$. The variance of the disturbance applied to each target, $\sigma_{fi}$ is also used to compute the ratio $r_i = \sigma_{si}/\sigma_{fi}$. A small $r_i$ can be used as evidence that the user is trying to select $x_i$. 

Williamson and Murray-Smith\cite{williamson2004pointing} define a threshold $v<1$ so that, when this evidence is strong, the probability that the user is trying to select $x_i$ increases, and decreases otherwise.
Starting from an even prior for all targets, when $r_i < v$, they keep track of an intermediary weight value for each target, which increases additively by $\alpha \times r_i$, and is attenuated multiplicatively by $\beta$ otherwise. These intermediate values are then normalised to produce the new probabilities and, when the resulting entropy falls below a threshold, the target $i$ with the highest $p(x_i)$ is selected. 
The choices of $\alpha$, $\beta$, and $v$ affect how quickly and how accurately the system decides that the user is making a selection. 

In comparison to deterministic approaches, Williamson and Murray-Smith's probabilistic approach has several advantages. The first is that it considers the probability of the targets jointly and provides a measure of uncertainty that can be used to indicate the most probable target when the uncertainty is low, i.e., when the entropy of the system is sufficiently small. The second is that it allows the evidence from the measurements to be integrated in a probabilistic update process as a function of prior probabilities. A third advantage compared to typical probabilistic gesture-based methods is that it neither requires training nor a specific shape model. Despite this early work, works in motion correlation since then have not used probabilistic techniques for inferring which target the user is trying to select.

In this work, we take direct inspiration from the pioneering work of Williamson and Murray-Smith, abstracting it to model later approaches and extending it to enable the analysis and design of new interfaces. Our work is framed within the larger program of \textit{computational interaction}, which ``focuses on the use of algorithms and mathematical models to explain and enhance interaction"~\cite{oulasvirta2018computational}. Core to computational interaction is mathematical modelling, typically involving ways for updating the model using data observed from the user~\cite{oulasvirta2018computational}. In this paper, we model motion correlation using a probabilistic approach, which enables us to derive a set of techniques for examining motion correlation designs before collecting user data.

\section{Probabilistic Modelling of the Selection Task}
Our thesis in this paper is that we can better understand the principles and make more informed and confident interface design decisions by looking at motion correlation interfaces from a probabilistic perspective. As such, our goal is to bring ideas from probabilistic interaction to researchers and practitioners working on the development of motion correlation interfaces.
In this and the next few sections, we offer a didactic primer on probabilistic input and demonstrate how previous motion correlation techniques can be formulated in probabilistic terms. As we discuss how to model deterministic techniques as probabilistic, we formalise a set of principles that are useful for the design of motion correlation interfaces.

In addition, we offer a practical example of how to employ the techniques we propose in the case study of \textit{Orbits}~\cite{esteves2015orbitsa}, a gaze-only motion correlation technique for interaction with smart watches. The design of \textit{Orbits} was done through an empirical process, collecting user data at each stage of the development. Here, we demonstrate the additional insights that are gained by analysing these interactions through a probabilistic framework.  

\begin{figure}[ht]
    \centering
    \includegraphics[width=1\columnwidth]{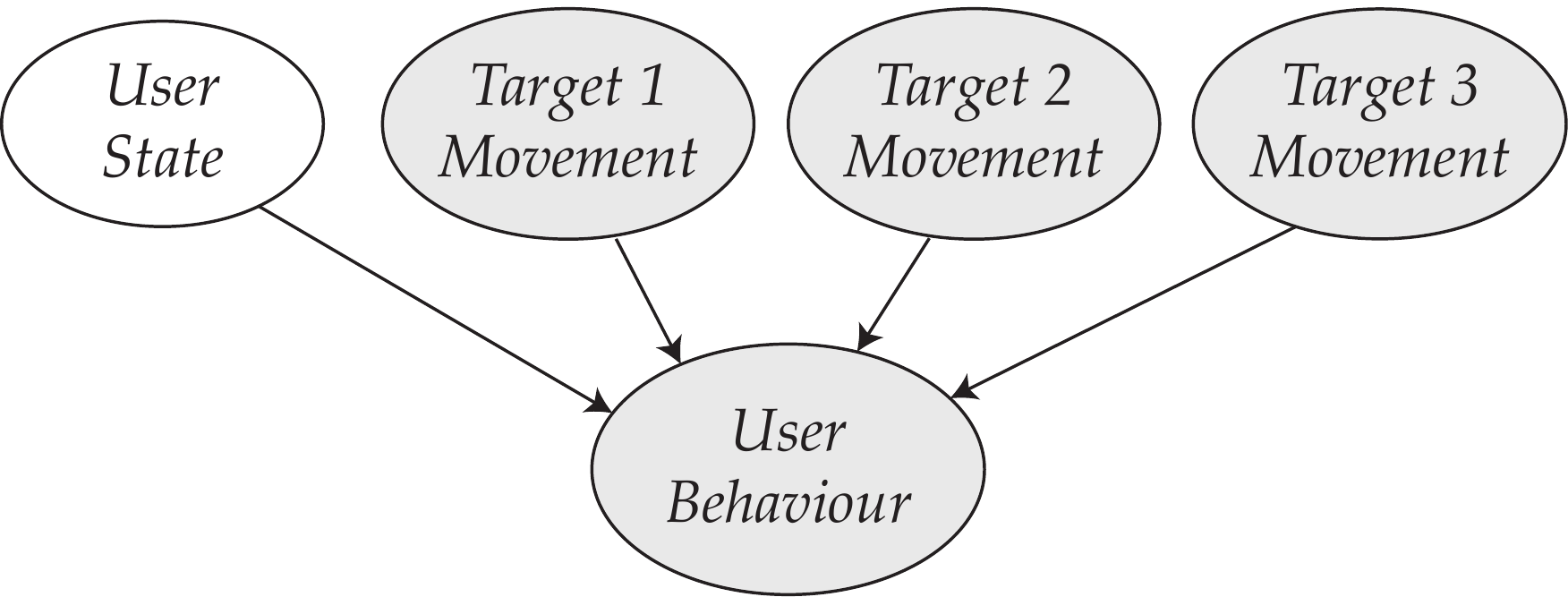}
    \caption{In a motion correlation interface the user movement is influenced by the user state and by the system state, represented here by each separate target's motion. Observed variables are in grey.}
    \label{fig:bayes1}
\end{figure}

To begin, we demonstrate how to formalise the motion correlation selection task in a notation that allows us to discuss it in probabilistic terms. The basic challenge in any selection task is to discover the user's state (e.g. the intention to select a target, if any) given a system state (e.g. a graphical user interface with moving targets) and a sequence of user behaviours (e.g. matching the motion of the intended target or natural behaviour when not attempting to make a selection). From a probabilistic perspective, we can model these as random variables, where the user behaviour is influenced by the user state and the system state. If we break the system state into separate random variables for each target, we can build the Bayesian network in Figure \ref{fig:bayes1}. 

More formally, the user state is a random variable with a sample space containing the intention to select each target ($x_i$) and the intention to not select any of them ($x_\emptyset$). 

\begin{equation}
    User\;State = X \in \left\{x_\emptyset, x_1, \ldots , x_N\right\}
\end{equation}

The target motions ($T_i$) and the user behaviour ($U$) are sequences of two-dimensional coordinates, not necessarily in the same coordinate system---e.g. the input device might not be calibrated to the output device. 

Like all Bayesian networks, the model in the figure encodes a joint distribution over the modelled variables. Missing edges reflect independence assumptions that are necessary to make probabilistic inferences efficient in runtime (computationally) and in observations needed for model fitting (statistically). In this case, we know that the variables $X, T_1, \ldots, T_N$ are pairwise independent by construction of the target motions. 

During the interaction, the target motions and the user behaviour are our \textit{evidence} variables (denoted in the figure in grey), as they are directly observable, whereas the user state is our \textit{query} variable (denoted in the figure in white), which we would like to infer. This is specified as the following conditional probability, which we can unpack with Bayes' rule. 

\begin{equation}
    P(X|U,T_1,\ldots,T_N) = \frac{P(U|X,T_1,\ldots,T_N)P(X)}{P(U,T_1,\ldots,T_N)}
\end{equation}

The key term in this equation is the likelihood $P(U|X,T_1,\ldots,T_N)$. This term computes for each possible user state, the probability of observing a given user behaviour given the motions available on the interface and the user state. A well-designed motion correlation interface displays motions that elicit sufficiently unusual user movements---decreasing $P(U|x_{\emptyset},T_1,\ldots,T_N)$---and that could only happen given that the motion is displayed on the interface--- increasing $P(U\approx T_i|x_i,T_1,\ldots,T_N)$, while balancing subjective user experience factors such as comfort, effort, and social acceptability. The challenge lies in how to compute this likelihood.

The motions of the targets should also be sufficiently different from each other, so that $P(U\approx T_j|x_{i\ne j},T_1,\ldots,T_N)$, that is, the likelihood of observing a user movement that similar to target $j$ when trying to select target $i$ should be small. Such difference can be created a priori by through the path design or in real-time, such as in Williamson and Murray-Smith's decorrelation approach~\cite{williamson2004pointing}. In their approach, as the probability of a target increases, so does the range of the disturbances on the target's motion characteristics (e.g. path, speed, etc.), to make the motion of the target even more distinct than the others, as if testing the hypothesis that the user is indeed trying to select it.  

With this framing, we can introduce a formal notation to known principles of motion correlation design. 

\begin{tcolorbox}[colback=white!80!gray,
colframe=white, fonttitle=\bfseries,
colbacktitle=white!50!gray,
coltitle=black,title=Principle 1: Motions in the interface should elicit a similar movement from the user]
We aim to maximise $P(U\approx T_i|x_i,T_1,\ldots,T_N)$---the likelihood that given a set of target motions and given that the user is trying to select target $i$, the movement observed from the user should be similar to the motion of target $i$.
\end{tcolorbox}

\begin{tcolorbox}[colback=white!80!gray,
colframe=white, fonttitle=\bfseries,
colbacktitle=white!50!gray,
coltitle=black,title=Principle 2: Targets should elicit motions distinct from non-interaction behaviour]
The motions of the targets on a motion correlation interface should not elicit movements that frequently occur in the user's natural behaviour when they are not interacting with the system. Formally, we aim to minimise $P(U\approx T_i|x_{\emptyset},T_1,\ldots,T_N)$---the likelihood of observing a user movement similar to a target motion when the user is not interacting with the system.
\end{tcolorbox}

\begin{tcolorbox}[colback=white!80!gray,
colframe=white, fonttitle=\bfseries,
colbacktitle=white!50!gray,
coltitle=black,title=Principle 3: Targets should display motions distinct from each other]
We aim to minimise $P(U\approx T_j|x_{i\ne j},T_1,\ldots,T_N)$---the likelihood of observing a user movement similar to target $j$ when the user is trying to select a different target $i$.
\end{tcolorbox}

\subsection{Case Study: Modelling Orbits}
In their evaluation of the performance of \textit{Orbits}, Esteves et al. measured the effects of the number of targets (2, 4, 8, 16), trajectory size (4.25$^o$, 2.62$^o$, and 0.98$^o$) and target speed (120$^o/s$, 180$^o/s$, and 240$^o/s$) on the true- and false-positives in a selection task~\cite{esteves2015orbitsb}. In all conditions, they used a fixed window size of 1s, a 30Hz eye tracker, and a Pearson's correlation threshold of 0.8. The targets moved along a circular trajectory, with half of the targets moving in the opposite direction of the other half. This is representative of the types of evaluation procedures found in previous works. 

In the case of \textit{Orbits}, the user state is the desire to select a target or not, and their behaviour is captured by the system as $X,Y$ coordinates provided by the eye tracker. Because the user most often looks straight at the screen, it is expected that when following a target, the user's eyes will follow a smooth pursuit movement that closely resembles the motion of the target (Principle 1). The head-mounted eye tracker employed in \textit{Orbits} captures data even when the user is not interacting with the system, so we must consider that the $x_{\emptyset}$ state will be over-represented---that is, most of the time, the user will not be interacting with the system. However, because the eyes only engage in smooth pursuits when there is a moving target for them to track, if the system is able to accurately detect whether the eyes are in this state, Principle 2 should be satisfied. The targets move around circular trajectories. Differently from shapes with straight edges, no two windows in a circle are the same, so all targets will exhibit a different motion at all points in time (satisfying Principle 3).

\section{Interaction as Transmission of Information}
The discussion above emphasises that ultimately, we seek to compute likelihoods. The novel insight in the motion correlation literature was that we can estimate these likelihood from the similarity between the motion of the target and the movement that user is making as captured by the input device. We can model this interaction from an information theory perspective, as \textit{transmission of information}~\cite{hornbaek2017whatisinteraction}. From this perspective, when the user is attempting to select a target, that target motion is like a message passing through a noisy channel and manifesting itself as the user behaviour. As such, the measurements $U(t)$ can be seen as a distorted version of the selected target trajectory (or none). These distortions might include noise, delays, and spatial transformations introduced by the sensors, processing hardware, geometric setup, and user movement inaccuracies. We represent these distortions as a spatial transformation matrix $A$, a time delay $\tau$, and a noise component $\epsilon$. Note that here we assume linear distortions, where distortions found in the real world, such as those introduced by camera lenses, might not be linear. More complex distortions require more sophisticated modelling approaches or different assumptions made by the interaction technique.

\begin{equation}
    X=x_i \implies U(t) = A \cdot T_i(t - \tau) + \epsilon(t)
    \label{eq:noise}
\end{equation}

By treating the interaction process as transmission of information, we can detect the evidence of whether a selection is taking place by measuring how much information is being transmitted. The \textit{mutual information}---the amount of information you gain about one variable by learning the value of another~\cite{mackay2003information}---can then be used as evidence that the interaction is taking place and that a selection is being attempted. This consolidates Principles 1--3 into a unified principle:

\begin{tcolorbox}[colback=white!80!gray,
colframe=white, fonttitle=\bfseries,
colbacktitle=white!50!gray,
coltitle=black,title=Principle 4: The intent to select is evidenced by the amount of mutual information between the target and user behaviours]
$P(x_i) \propto I(U;T_i)$---that is, the higher the mutual information between the motions of the user and the target, the higher the likelihood that the user intends to select that target.
\end{tcolorbox}

\begin{figure}[ht]
    \centering
    \includegraphics[width=0.8\columnwidth]{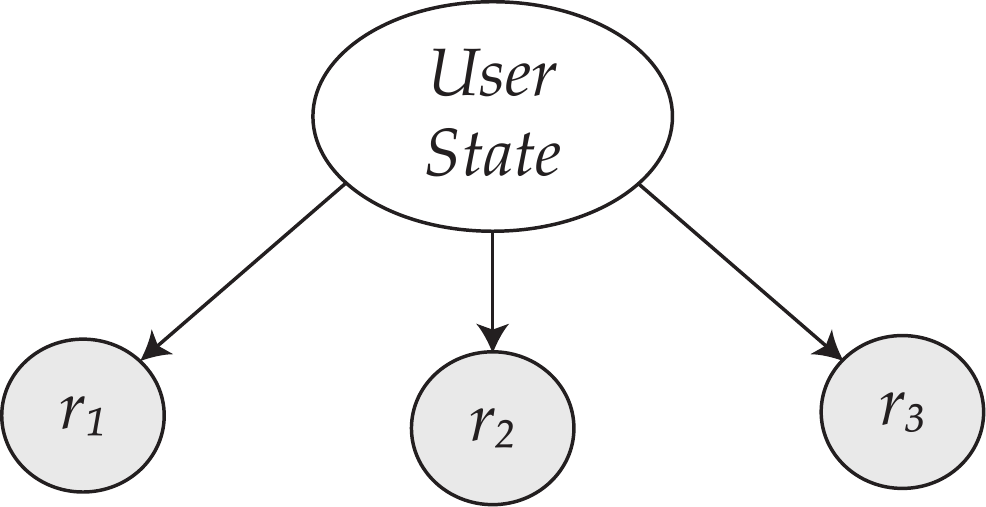}
    \caption{By measuring the similarity between the user movement and the motion of each target, we can more easily compute the probability that the user is in a given state. Observed variables are in grey.}
    \label{fig:bayes2}
\end{figure}

The insight that mutual information is the key idea behind the selection means that the term ``correlation'' in ``motion correlation'' should be understood in the broad English sense of the word rather than as Pearson's correlation coefficient. In practice, however, computing the mutual information is challenging~\cite{walters2009mutualinformation}, so we must estimate it using other \textit{similarity measures}, such as Pearson's product-moment correlation coefficient, as used in Vidal et al.'s \textit{Pursuits}~\cite{vidal2013pursuitsb}. We denote this similarity measure as $r(U,T_i) = r_i$. Therefore, instead of computing $P(X|U,T_1,\ldots,T_N)$, we can compute $P(X|r_1,\ldots,r_N)$, which is much easier. Our network then becomes the one shown in Figure \ref{fig:bayes2}.

Besides Pearson's coefficient, 
many other similarity functions have been used in previous work. We have already described how 
Williamson and Murray-Smith
used a ratio of variances as their similarity function~\cite{williamson2004pointing}. 
Fekete et al. proposed several measures computed on the path derivatives, including the Euclidean distance, the normalised Euclidean distance, and Pearson's correlation, ultimately opting for the normalised Euclidean distance~\cite{fekete2009motion}. 
Carter et al. also used Pearson's correlation, but prior to computing the coefficient, they rotated the data to ensure an even distribution of variance between the horizontal and vertical axes~\cite{carter2016pathsync}. Velloso et al. proposed a version of Pearson's correlation that considered the variance in both axes simultaneously~\cite{velloso2018circular}. Drewes et al. used the linear regression slope~\cite{drewes2018dialplates}. Zhang et al. compared the dominant frequencies of the signal~\cite{zhang2017fingorbits}.

The choice of similarity measure depends on the assumptions about the data. For example, if we assume no calibration between the input device and the display---i.e. we do not know the transformation matrix $A$---it becomes difficult to find appropriate thresholds for using the linear regression slope or anything based on Euclidean distances. Though measures based on Pearson's correlation work well in these cases, they are susceptible to statistical factors known to affect this coefficient, such as the distributions of the data being compared, the variability of the data, the lack of linearity, the presence of outliers, etc.---for a review, see Goodwin and Leech~\cite{goodwin2006correlation}. Among them, one that is of particular relevance to the design of motion correlation interfaces is the fact that it expects a similar distribution (preferably normal) in the samples being compared. This is difficult to guarantee with regular shapes, such as the square, where there might be windows with no variation in one of the coordinates when the target is traversing one of the edges. The correlation coefficient is also susceptible to the presence of outliers, and this effect is more pronounced in small windows. As a rule of thumb, Drewes et al. suggest that the window size should be set to allow the motion of the target by 3-4 times the amplitude of the measurement noise~\cite{drewes2018speedsandtrajectories}. The choice of measure also depends on which properties characterise the differences between the target motions. Certain measures, like Zhang et al.'s frequency analysis~\cite{zhang2017fingorbits} work best when targets display motions with different \textit{frequencies}, whereas others, like Pearson's correlation, work best when the targets display motions with different \textit{phases}, but the same frequency. 

Most systems assume a sufficiently small lag $\tau$ and error $\epsilon(t)$, and a stable calibration matrix $A$. However, these can be measured in pilot tests and incorporated into the calculation of the similarity, either by shifting the windows being compared (in the case of a substantial $\tau$) or by filtering the input data (in the case of a substantial $\epsilon(t)$).

\begin{tcolorbox}[colback=white!80!gray,
colframe=white, fonttitle=\bfseries,
colbacktitle=white!50!gray,
coltitle=black,title=Principle 5: The appropriateness of a similarity measure is determined by how well it estimates the mutual information]
Similarity measures should be chosen based on the possible distortions of the channel and how much information about the user state can be determined from the knowledge of the target states.
\end{tcolorbox}

\subsection{Case Study: Orbits as Transmission of Information}

Because \textit{Orbits} was designed for interaction with smart watches while the user is wearing a head-mounted eye tracker, we cannot assume that the gaze data will be in the same coordinate system as the GUI in the watch. This means that the matrix $A$ in Equation \ref{eq:noise} will contain a scaling factor and that the noise component $\epsilon(t)$ will have a constant translation offset together with a random component. We assume that the two devices will be synchronised, making $\tau$ sufficiently small to be ignored. The user's head and the watch might not be perfectly aligned. This is a problem, particularly in the case of targets moving around a circle, as the rotated image will lead to ambiguity. This can either be fixed by using the scene camera of the eye tracker/the IMU in the watch to re-align them or by assuming that this alignment is a pre-condition for the interaction, so we will not assume that $A$ contains a rotation component. These assumptions mean that certain similarity measures cannot be applied. Ones based on Euclidean distance cannot be used because they are not invariant to translation and scale~\cite{fekete2009motion}; one based on the linear regression slope cannot be applied because it is not invariant to scale~\cite{drewes2018dialplates}; and one based on frequency analysis would fail due to the fact that all targets move with the same frequency~\cite{zhang2017fingorbits}.

\section{Computing Likelihoods}
Previous works typically make a decision about which target to select deterministically, by simply thresholding the similarity at an empirically obtained value $\lambda$. We can model these approaches, by using the following conditional probability density functions.


\begin{equation}
 p(x_i|r_i) =  
    \begin{cases}
     1 & \text{if}\ r_i>\lambda,\\
      0 & \text{otherwise}
    \end{cases}
\end{equation}

\begin{equation}
p(x_\emptyset|r_i) = 1 - \sum_i^Np(x_i|r_i)
\end{equation}

This formulation already exposes a few problems with this approach. First, it does not deal well with cases where more than one target cross the threshold---all targets that cross get assigned an equal likelihood of 1. In practical implementations, this is typically solved by only selecting the target with the highest similarity. Nevertheless, it would be good to be able to incorporate this ambiguity into the decision to whether make a selection and provide some form of user feedback, reflecting in the interface the uncertainty in the system (for example, Williamson and Murray-Smith showed a growing circle around the target as its likelihood increased in relation to the other targets~\cite{williamson2004pointing} and Schwarz et al. delayed system actions that were difficult to reverse until the system was confident enough of the user intent~\cite{schwarz2010framework}).

The second insight is that the step function lacks nuance. Particularly for values near the threshold, this can cause the probability to alternate between 0 and 1 with small perturbations in the output of the similarity measure. A smoother thresholding function could help address this problem and also assign higher probabilities to targets with a higher similarity. A simple solution is to use a probability density function with a soft threshold, such as the probit or logit distributions. These approaches do not suffer from the discontinuity at the threshold, smoothly increasing its outputs as the likelihood of the input increases. The downside of these approaches is that they still require the system designer to manually specify a mean threshold. 

A potentially better approach is to collect user data to learn these distributions. To do so, we must collect data of users attempting to select targets, along with the target motion data in order to compute the similarity between them. Likewise, we must also collect data of users not trying to select any target, but rather, performing tasks appropriate for the context of the system. 

For example, a video-on-demand application as featured in \textit{AmbiGaze} should consider collecting data not only of users attempting to control the interface widgets, but also of users watching a variety of videos~\cite{velloso2016ambigaze}. One can then simulate target motions to compute the similarity values against the natural behaviour data. The system can then compute the similarity between the gaze behaviour and the true target and estimate a probability density function $p(r_i|x_i)$, where $x_i$ is the true state. The other distributions $p(r_i|x_j)$, where $i\ne j$, can be computed by modifying the gaze or trajectory data (e.g. if the target motions differ in phase, this can be simulated by shifting the data). Finally, the trajectory data of the target motion can be compared to the natural behaviour dataset to generate the distribution $p(r_i|x_\emptyset)$.

By applying Bayes' rule, we notice a few more issues:
\begin{equation}
    P(x_i|r_1,\ldots,r_N) = \frac{P(r_i|x_i, r1,\ldots,r_{i-1},r_{i+1},\ldots,r_N)P(x_i)}{P(r_1,\ldots,r_N)}
\end{equation}

 In order to obtain the step function used in deterministic approaches, the underlying assumption in these works is that when the user is following target i, $P(r_i|x_i, r_1,\ldots,r_{i-1},r_{i+1},\ldots,r_N) = 1$, and it equals zero otherwise. This implies that it is impossible to observe a correlation above the threshold if the user is not following the target and impossible to observe a correlation below the threshold when they are. Clearly this is a very strong assumption, easily disproven with empirical tests. The consequence is that the system might make very confident decisions based on flawed assumptions, leading to a higher error rate.

This formulation also reveals another issue with previous approaches. Typically, they compare each target individually and threshold the similarity without taking into account the similarity with other targets. In practice, this means that these techniques consider the following to be true
\begin{equation}
    P(r_i|x_i, r_1,\ldots,r_{i-1},r_{i+1},\ldots,r_N) = P(r_i|x_i)
\end{equation}
This is not necessarily a problem: by making the similarities conditionally independent given $x_i$, we reduce the complexity of the inference---this is the assumption that all Na\"{i}ve Bayes models make. However, $r_i$ are not necessarily independent: if the motion of a target $T_1$ is correlated with the motion of target $T_2$ and $T_2$ is correlated with $T_3$, most likely $T_1$ is also correlated with $T_3$~\footnote{It is important to note that $T_1$ and $T_3$ will only be positively correlated if $r(T_1,T2)$ and $r(T_2,T_3)$ are sufficiently close to 1, but not necessarily true otherwise, as demonstrated by Langford et al.~\cite{langford2001correlationtransitivity}---in other words, being positively correlated is not a transitive property.}.  The problem is that we often \textit{only} consider $P(r_i|x_i)$ when computing the likelihood, when in fact all other $P(r_j|x_i)$ also give us evidence for the probability that $P(x_i|r_1,\ldots,r_N)$.



Shifting our attention to the other terms in the equation, the denominator $P(r_1,\ldots,r_N)$ is just a normalisation factor to make the probabilities add up to 1. The prior probability $P(x_i)$, on the other hand, can open up opportunities for tuning the algorithm. A na\"{i}ve---though standard---approach would be to set all prior probabilities to be the same, effectively making this term irrelevant for the state estimation. This is the approach underlying current deterministic approaches, even though there is no real reason to assume that all prior probabilities would be the same in every case. The beauty of probabilistic input is precisely in enabling developers to incorporate prior knowledge in a quantifiable manner.

Another approach would be to use domain knowledge or separate classifiers to estimate reasonable values. For example, a motion correlation interface based on smooth pursuits might use a separate classifier to identify when the eyes engage in a smooth pursuit and use its output as a prior. Alternatively, a gesture recognition algorithm could be used to determine the probability that the gesture has the right shape before trying to determine which target along that shape is being followed.  Yet another approach would be to collect statistical data of application usage to compute these values. This is particularly valuable in cases where users exhibit consistent usage patterns, such as when typing or navigating hierarchical menus. A final possibility is using the probabilities computed at previous points in time to adjust our confidence over time. From this perspective, the Bayesian networks we presented so far can be extended into temporal inference models, leveraging techniques such as Kalman filters~\cite{Biswas:2013} and particle filters~\cite{rogers2010fingercloud}.

\subsection{Case Study: Computing likelihood distributions}
Esteves et al. used a single similarity threshold $\lambda = .8$ to determine a selection. As we discuss above, this approach has several limitations. To demonstrate a probabilistic approach based on empirical likelihoods, we analysed the dataset collected as part of the development of \textit{Orbits}~\cite{esteves2015orbitsa}. This dataset contained data of users following circular targets with their eyes, as well as other natural gaze data, including reading the time, watching a video, playing a video game, and reading a news article. The gaze data was sampled at 30Hz and compared in windows of 30 samples. 

\begin{figure}[t]
    \centering
    \includegraphics[width=1\columnwidth]{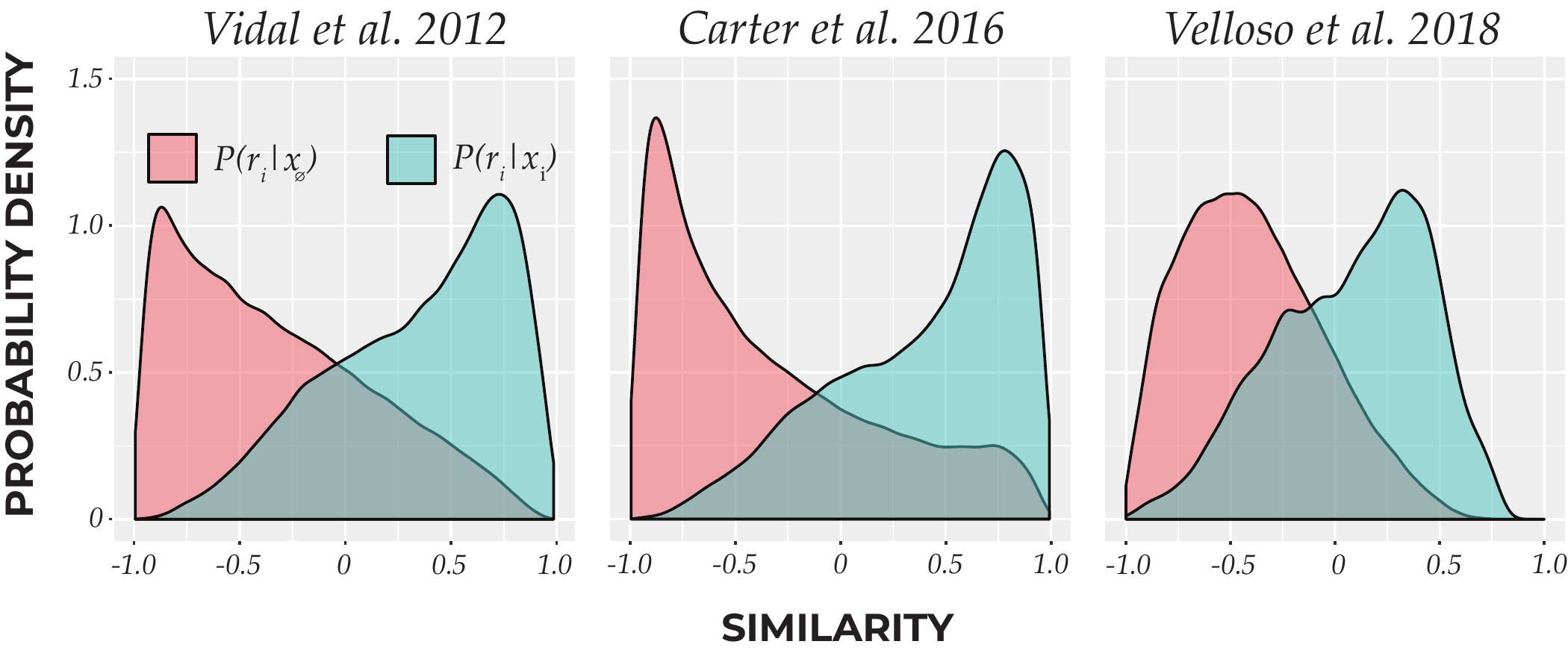}
    \caption{Probability density functions for three similarity measures found in the literature. Note that all measures yield distributions that overlap and are therefore substantially different to the step function assumed by deterministic methods.}
    \label{fig:pdfs}
\end{figure}

We computed the similarity between the gaze data and the target using three metrics: the Pearson's correlation coefficient of the lowest axis proposed by Vidal et al. ~\cite{vidal2013pursuitsb}, the modification of this measure involving a rotation of the windows to maximise the variance across axes proposed by Carter et al.~\cite{carter2016pathsync}, and the complement of the ratio of the sum of squares computed in two dimensions proposed by Velloso et al.~\cite{velloso2018circular}. In the cases where the user was following a target ($X=x_i$), the similarity was computed between the gaze data and the target positions. In the case where the user was not following any target ($X=x_\emptyset$), we simulated a target using the same characteristics of the targets in the positive condition. Figure~\ref{fig:pdfs} shows the probability density functions $P(r_i|x_i)$ and $P(r_i|x_\emptyset)$, computed using a Gaussian kernel density estimate. As predicted, the empirical probability density functions have a much softer distribution compared to the step function assumed in deterministic approaches. These curves can then be used to choose an appropriate threshold that gives more weight to robustness or responsiveness depending on the requirements of the system.

\section{Considering all targets simultaneously}
In the previous section, we discussed how prior approaches only take into account the similarity between the target motion and the projection of the user movement onto the input space when computing the likelihood that the user is trying to select that target.  However, even if the user is following the target perfectly, without any distortion (i.e. $U = T_i$) this motion can nevertheless  present a degree of positive correlation with the other targets, especially if the trajectories are not carefully designed.  Ideally, if the true state is $x_i$, $r(U,T_i)$ should be much higher than $r(U,T_j)$ for all $j\neq i$. However, depending on the choice of similarity measure, trajectory shapes, and window sizes, this is not always achievable. Therefore, understanding the expected levels of similarity between $U$ and all $T_j$ when the user is following $T_i$ can help us make decisions about the quality of the trajectory design.

\begin{figure}[t]
    \centering
    \includegraphics[width=0.7\columnwidth]{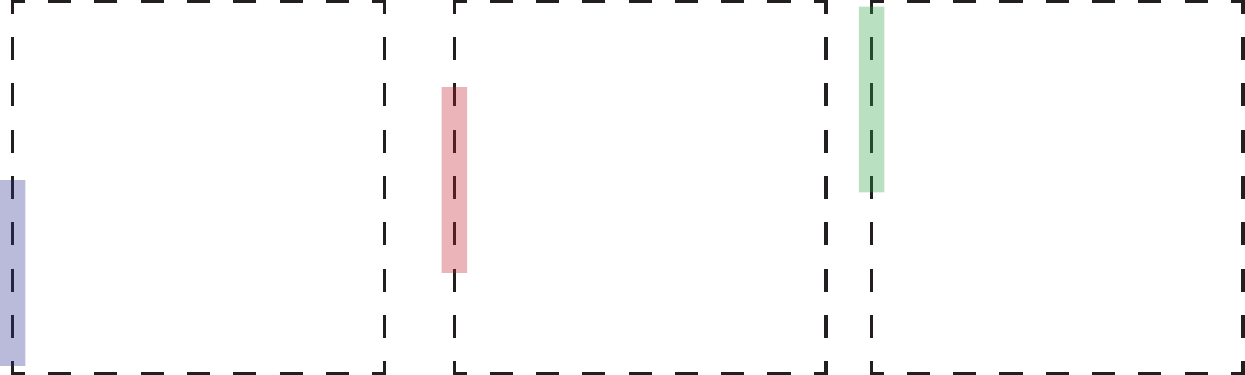}
    \caption{When the path has straight edges that are longer than the distance traversed by the target within the window, multiple trajectory windows will look the same, making it impossible to distinguish between them. In the example, we see three targets traversing the edge of a square path one after another. Even though they are temporally offset, their relative trajectories will look the same until they reach the corner.}

    \label{fig:similar_windows}
\end{figure}

This problem is exacerbated in the case of polygons where distance traversed by the targets within a window is smaller than the length of the edges, such as the one in Figure \ref{fig:similar_windows}. The figure shows a square where targets traverse half the length of the edge within a window and all three targets move in the same direction. The trajectories in all three windows (as well as all other windows between them) look exactly the same in terms of their relative motion. If a calibration-free system observes a straight vertical line from the input device, it will know that the user is following \textit{one} of them, but not \textit{which} of them. In this case, the system would have to wait until the targets start hitting the corner of the polygon to be able to tell them apart. From an information theory perspective, the corner of the square has more information than the edge, making it more distinguishable. 

\begin{figure*}[t]
    \centering
    \includegraphics[width=2\columnwidth]{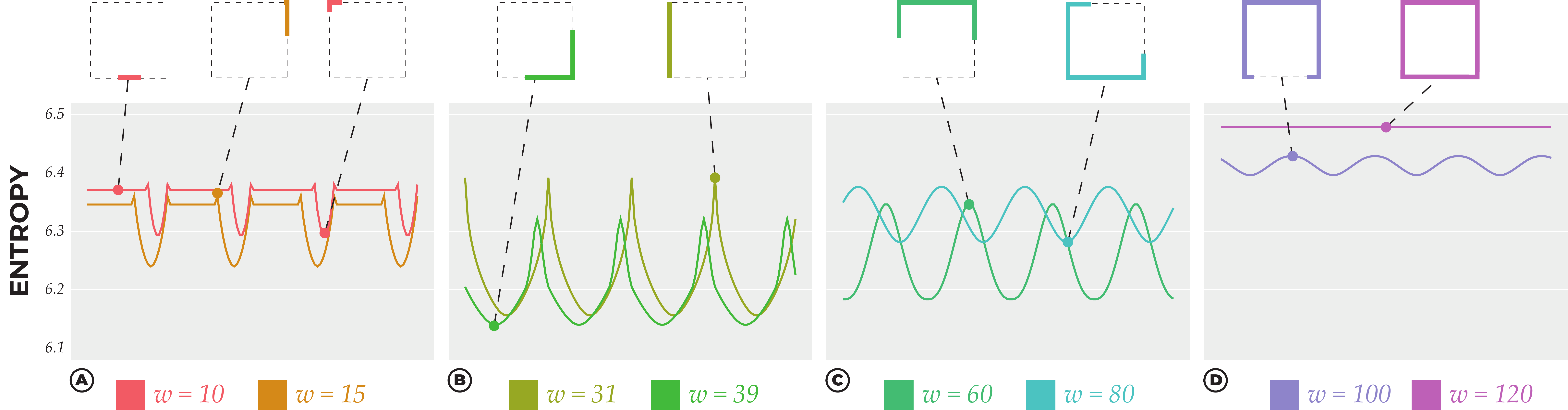}
    \caption{The window size affects the overall entropy of the system. Different windows have different amounts of information, leading to varying levels of entropy. The x-axis represents the index of the window. Increasing the window size does not necessarily lead to a smaller entropy.}
    \label{fig:entropy_w}
\end{figure*}

In practice, the differences in the information content of the windows of a trajectory mean that the point in time where the user begins to follow a target matters for the interaction---\textit{when ideally it should not}: a target that is just entering the edge of the square will be harder to select than one that is about to hit the corner. This might affect the interaction by delaying the selection decision. If we are able to quantify how distinct any given window in a trajectory is to the others, we can have an initial estimate of the quality of that trajectory for a motion correlation interface. The key idea here is that motion correlation interfaces must not only consider the targets trajectories (i.e. space + time), but also the path itself (i.e. space alone).

As such, we propose the use of entropy for understanding how  distinguishable each window along a path is from the rest. For each window, we compute the similarity between that window and every other window and normalise these values to obtain probabilities. We then compute the entropy for that window. This measures how similar this window is to all others---a low value of entropy means that there are few windows that capture most of the high similarity values, which is what we want. We finally compute the entropy for all windows of the trajectory to see how it fluctuates as the target moves. The idea here is that by looking at the path itself, comparing the possible windows sampled from it to each other (as opposed to comparing them to windows sampled from the input), we can estimate the upper bound of the confidence that the system can have at a given point in time (as the path itself would correspond to perfect motion matching with no noise).

To demonstrate this technique, we generated a square trajectory with 120 points. We then computed the entropy for all possible windows along the trajectory, varying the window size---this is equivalent to having 120 targets moving along a square while the user follows one of them with no distortion. We used the Pearson's correlation of the most dissimilar axis~\cite{vidal2013pursuitsb,esteves2015orbitsa} as the similarity measure, and we repeated this process for windows ranging from 5 to 120 samples. In total, we computed $116$ (window sizes) $\times 120$ (windows on the path)  values of entropy, calculated based on $120$ comparisons between windows for each entropy value.

\begin{figure}[t]
    \centering
    \includegraphics[width=0.7\columnwidth]{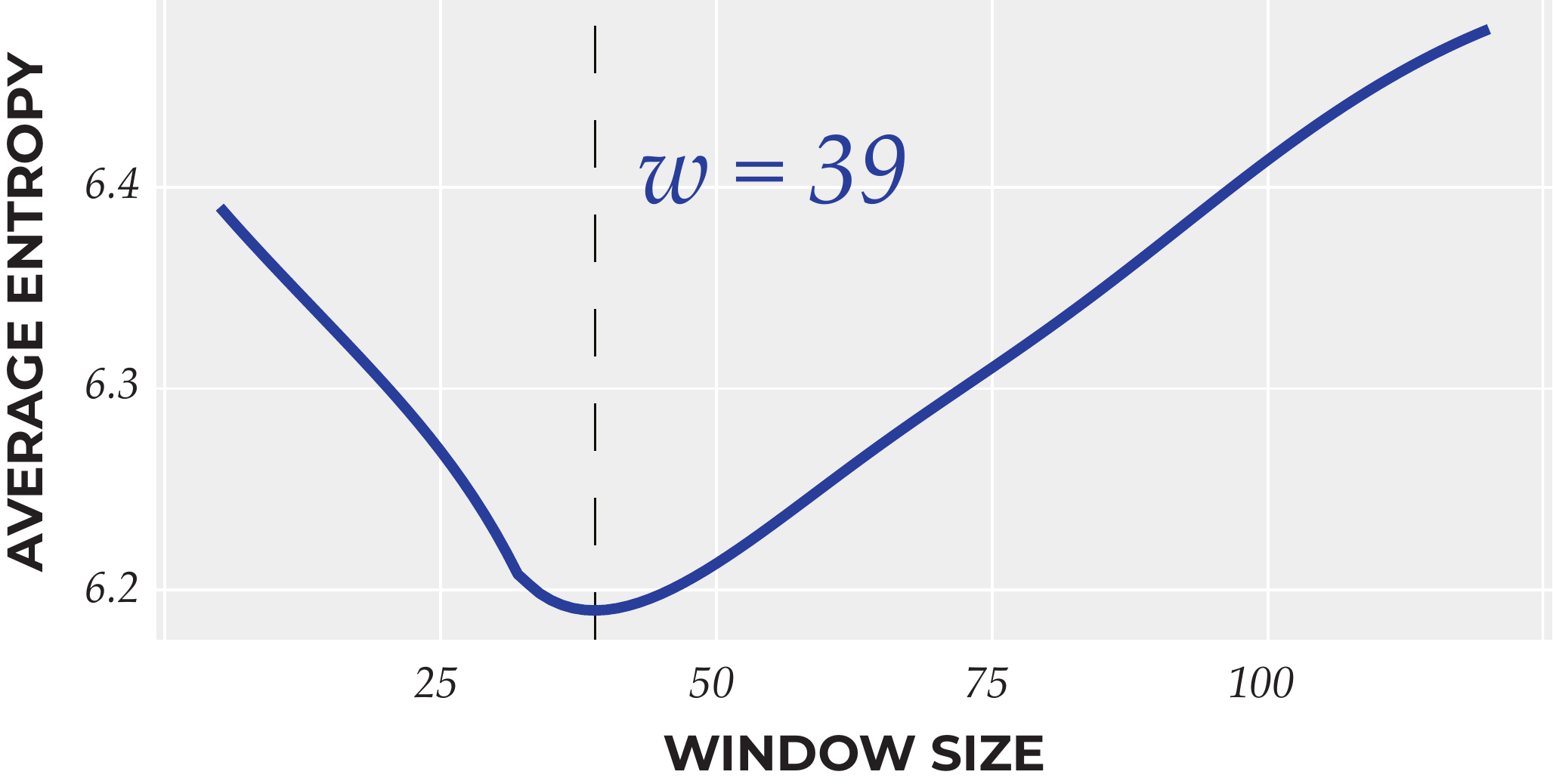}
    \caption{Average entropy across all windows as a function of the window size, for a square with 120 samples. A window size of approximately 1/3 of the perimeter minimises the entropy.}
    \label{fig:optimal_window}
\end{figure}

Figures \ref{fig:entropy_w} and  \ref{fig:optimal_window} show how the entropy varies along the square for different window sizes. In Figure \ref{fig:entropy_w}-A, the window sizes are shorter than the size of the edge, so plateaus of high entropy alternate with entropy drops at the windows that include the corners, with the lowest values at the windows centred at each corner. Because the x-axis refers to the first element of the window, as we increase the window sizes, the curves shift to the left. Also, as we increase the window size, the plateaus become shorter until they disappear once the window size becomes larger than the edge (Figure  \ref{fig:entropy_w}-B). The overall entropy hits a minimum when it is approximately one third of the perimeter of the square and begins to increase again. As it increases, it becomes sinusoidal (Figure \ref{fig:entropy_w}-C) with higher baselines and smaller amplitudes until it becomes a flat horizontal line when the window size is 100\% of the perimeter (Figure \ref{fig:entropy_w}-D).

Figure \ref{fig:optimal_window} shows how the average entropy across the whole trajectory varies with the window size---each point in the curve is the average of each curve like the ones in Figure \ref{fig:entropy_w}. In this case, the optimal window size would be 1/3 of the perimeter, but this will depend on the similarity measure and trajectory shape. Naturally, this window size is only optimal in terms of how well it distinguishes a window from the other windows in the shape, but in practice the optimal window will also depend on other factors, such as the interaction latency and robustness to noise. Nevertheless, this analysis can help us reduce the design space of possible window sizes to explore in empirical studies. Further, it allows us to quantify the effect of the starting point of the interaction on the expected performance of the system.

The previous analysis shows that the entropy for each window in the shape is  affected by the amount of information in each segment of the shape. An analysis like this can show that a priori, the system will have a higher degree of confusion at the edges than at the corners, so the developer can choose to defer a decision until it reaches the corner or change the system parameters (e.g. adjusting the window size to always include a corner). Further, the variations in entropy are not simply due to the shape of the trajectory---the measure itself leads to variations in performance along the same shape. This is best demonstrated by how the same pairs of input and target motion windows can yield different similarity values depending on the coordinate system (see a concrete example in the case study below).

\subsection{Case Study: Analysing the similarity between windows in Orbits}

In \textit{Orbits}, because targets move around circular trajectories, no two windows will look exactly the same, regardless of the window size, avoiding the problem shown in Figure \ref{fig:similar_windows}. However, depending on the point along the trajectory through which the target is passing and how far apart they are, they might nevertheless exhibit high correlation. 

\begin{figure}[t]
    \centering
    \includegraphics[width=1\columnwidth]{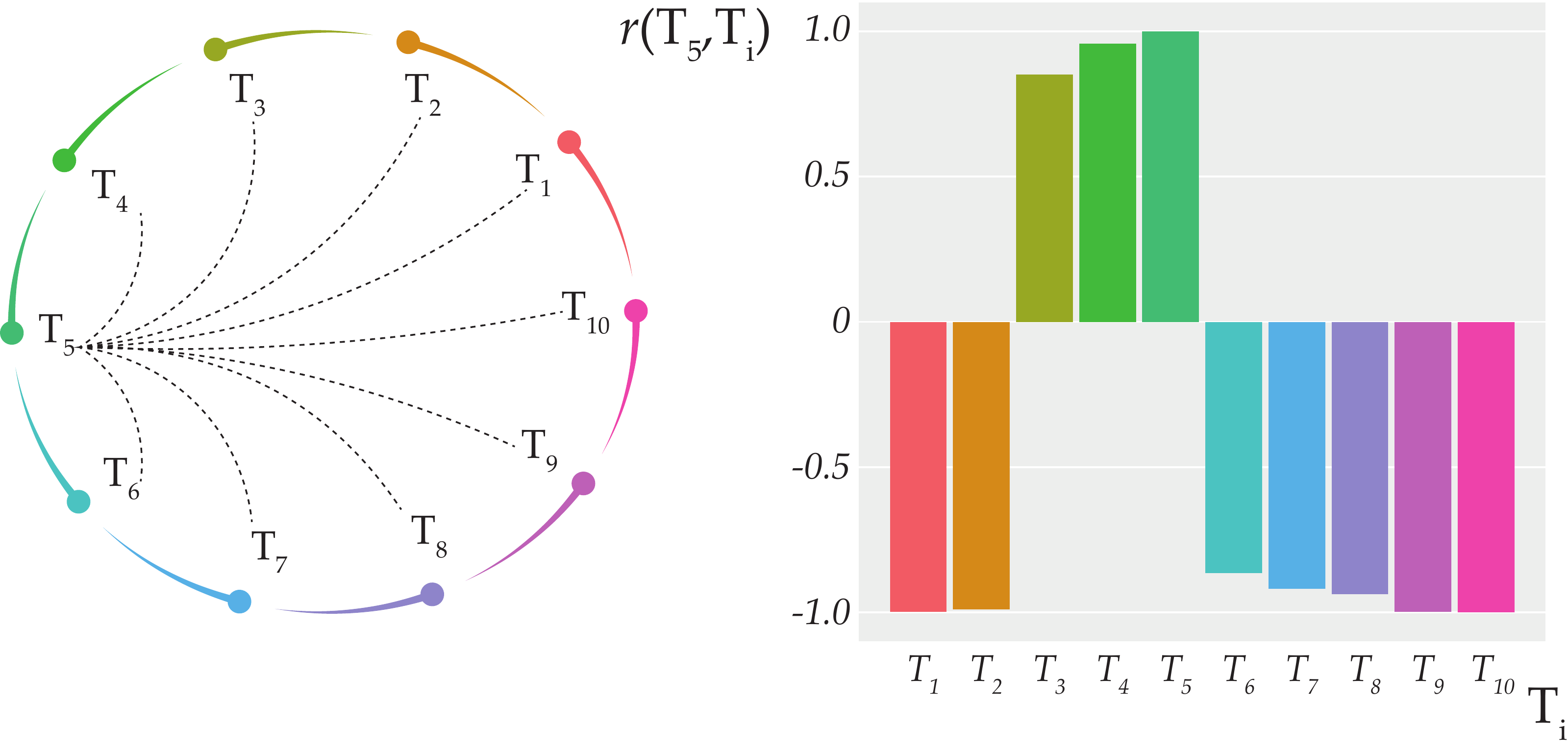}
    \caption{Similarity between all combinations of target $T_5$ and every possible target $T_i$ within the same window. Even if the user managed to perfectly follow the target, they would still yield a high similarity with other targets.}
    \label{fig:considering_all_windows}
\end{figure}

The easiest way to reveal this is to make pairwise comparisons between the target windows themselves (in contrast to the usual approach of comparing the targets' and the user's behaviours).
Figure \ref{fig:considering_all_windows} shows an example of this type of analysis in the context of \textit{Orbits}. Consider 10 targets moving around a circle with 160 samples. The figure shows a comparison between one arbitrary window ($w=15$) and all other windows using Vidal et al.'s method based on Pearson's correlation~\cite{vidal2013pursuitsb}. By comparing the targets to each other, we analyse the system performance as if the user was able to perfectly match the desired target's motion, giving us an upper bound on the system performance. In the case of the figure, the reference target is $T_5$. As the bar chart shows, the similarity between the target and itself is perfect ($r(T_5,T_5) = 1.0$)---as we should expect---but the similarity to other targets is also rather large: $r(T_5,T_4) = 0.96$ and $r(T_5,T_3) = 0.85$. In practice, this means that \textit{even if the user follows the motion of the target perfectly, we will still observe a high correlation with other targets}.

The choice of similarity measure also affects how the entropy of the design fluctuates along a path. To demonstrate this in the context of \textit{Orbits}, we generated a circular trajectory (radius = 1) with 60 samples, from which we sampled a window with 20 samples. We then simulated user input by distorting this trajectory with a Gaussian noise (SD = .1). We then computed the similarity between the original window and its distorted version using 4 similarity metrics described in the literature~\cite{vidal2013pursuitsb,carter2016pathsync,drewes2018speedsandtrajectories,velloso2018circular}. We then rotated these windows (both the original and the distorted) by angles ranging from zero to 360 degrees, computing the similarity for each rotation.

\begin{figure}[t]
    \centering
    \includegraphics[width=1\columnwidth]{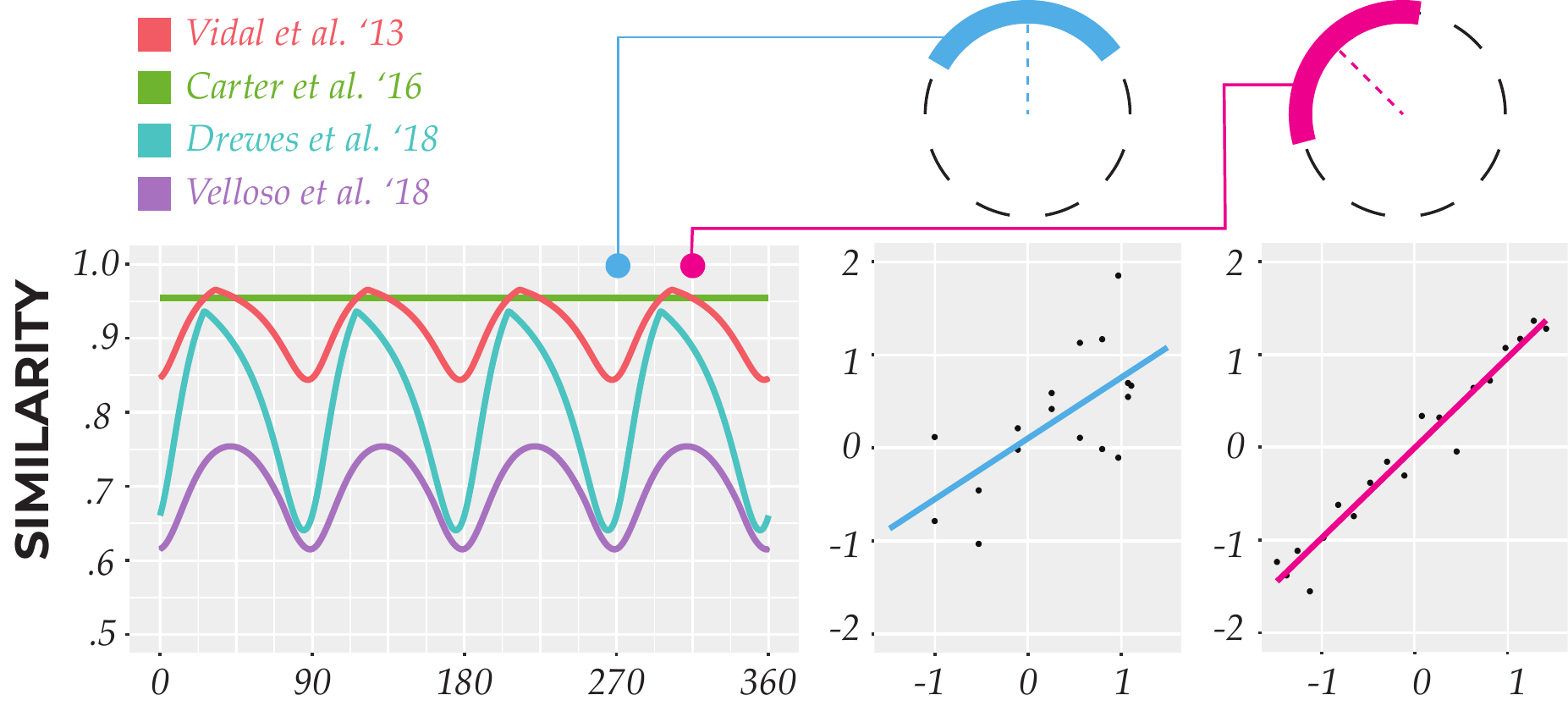}
    \caption{Left: The effect of the coordinate system on the output of different measures. Centre and Right: Correlation between the Y coordinate of the reference window and the noisy window for windows centred at North and Northwest points of the circle. When the variance in the data is more evenly distributed along each axis (pink), we observe a higher correlation than when it is concentrated in one of the axes (blue).}
    \label{fig:measure_vs_angle}
\end{figure}

Figure \ref{fig:measure_vs_angle}-Left shows our results for four different similarity measures. Though the absolute value of the similarity alone is not so important in this case---what matters is how well they are able to discriminate between correct and incorrect targets---there should not be much fluctuation in this value. This is because the data being compared is \textit{exactly the same}; the only difference is how the windows are orientated relative to the coordinate system. Notice that three of the measures are substantially affected by the orientation of the window. The plots in Figures \ref{fig:measure_vs_angle}-Centre and \ref{fig:measure_vs_angle}-Right help to illustrate why. In the case of the blue window, because there is little variance in the trajectory data in the Y axis the noise in the data ends up having a larger effect, leading to a lower correlation ($r=0.85$), as we can see in the middle plot. In the case of the pink window, both X and Y axes exhibit a more balanced variance, which in turn, leads to a more robust calculation of the correlation coefficient ($r=0.98$).
The only measure that is robust to these changes in rotation is Carter et al.'s \textit{Rotated Correlation}, which prior to calculating the correlation coefficient, rotates the window to better distribute the variance across the axes~\cite{carter2016pathsync}.

\section{Deciding when to make a selection}
In the previous section, we saw that current approaches are limited in how they handle multiple targets yielding high similarities, often just selecting the most similar one. Ideally, we should be able to quantify this uncertainty and make decisions about whether and which target to select based on our level of confidence considering the interface as a whole. Here, we draw directly from Williamson and Murray-Smith, who used a similarity metric to compute a weight for each potential target, normalised them into probabilities and computed the entropy of the system to decide whether to make a selection~\cite{williamson2004pointing,mackay2003information}. For us, entropy is interesting because it quantifies how concentrated the probabilities are around any of the targets.





The entropy value can give us a hint as to the number of states between which system is undecided: if the entropy $H(X) = \log_{2}K$, we are uncertain between approximately $K$ states. As we gather more evidence that a certain target $x_i$ is being followed, we increase its probability $p(x_i)$ and reduce the probabilities of the other states $\{x_j | j \ne i\}$. As $p(x_i)$ reaches one (and the others approach zero), the uncertainty is reduced and the entropy tends to zero. 
Therefore, this allows us to handle the case where multiple targets are above the threshold. As a few of them yield high likelihoods, the others will yield low ones. Looking at the probabilities individually would lead us to select one or all targets; looking at the entropy, we learn that the system is not confident enough in its estimation, so we might wait until more data is available before making a selection. In practice, we make a selection when the entropy of the system falls below a certain empirically determined threshold. Other simpler, but related approaches, such as the Gini coefficient or the ratio between the largest and second largest likelihoods can also be informative in this regard.

\begin{figure}
    \centering
    \includegraphics[width=1\columnwidth]{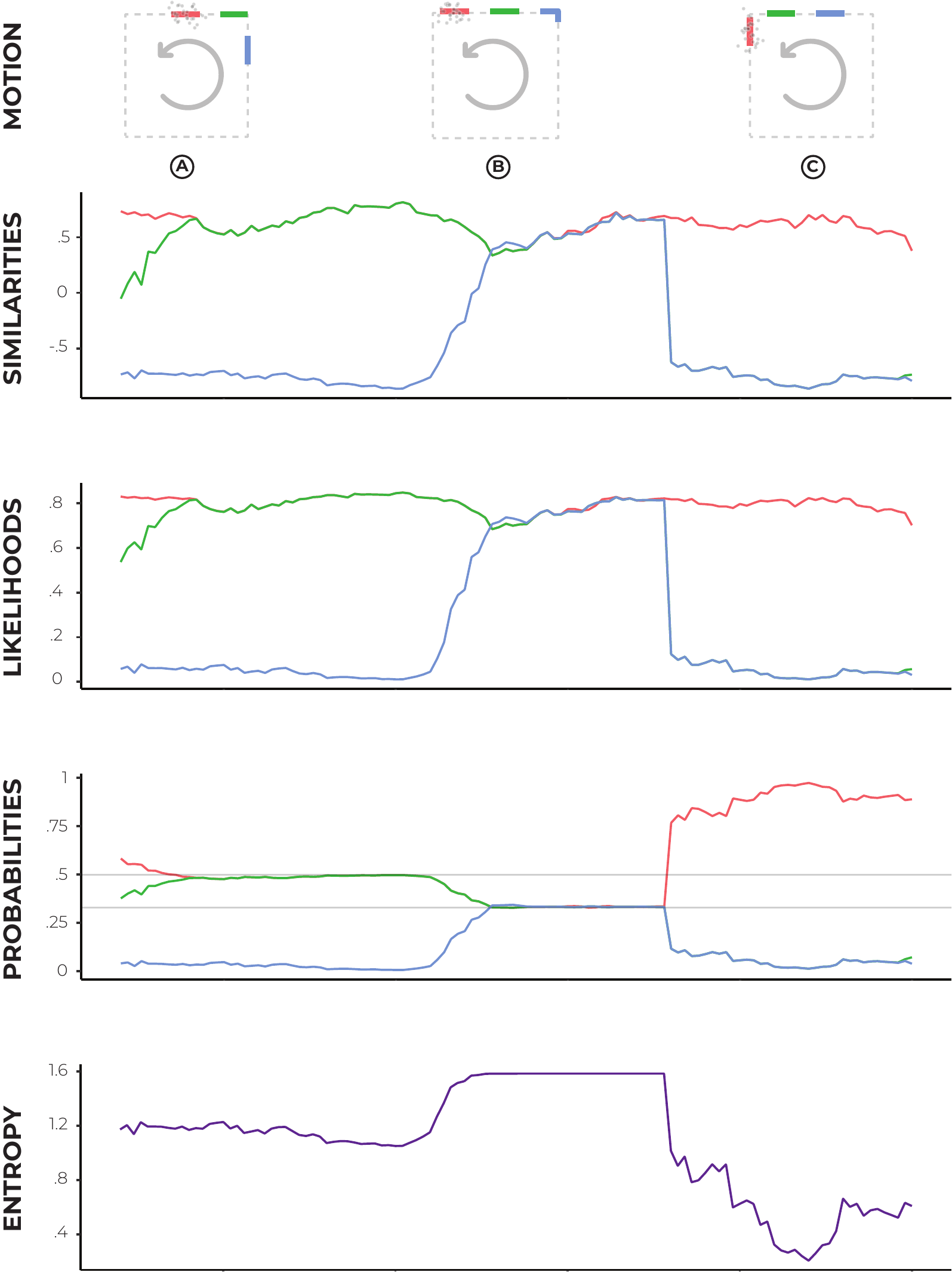}
    \caption{Stepping through a probabilistic calculation: Three targets move along a square path as the user noisily matches the motion of the red target. We compare the input signal to the trajectories of the targets, calculate the likelihoods using probability density functions, and normalise them to calculate the probabilities and the system entropy. Initially, two of the targets exhibit the same relative trajectory (A), yielding high similarities for both. When the third target reaches the edge (B), the entropy of the system increases. When the red target reaches the side edge of the square (C), its trajectory is sufficiently distinct from the others, leading to a drop in the entropy, which signals to the system that a selection can be made.}
    \label{fig:entropy_calculation}
\end{figure}

Figure \ref{fig:entropy_calculation} shows an example of how this works in practice in the case of three targets moving with constant speed around a square path, but slightly offset. The user data is shown in grey and corresponds to a noisy version of the red trajectory. We computed the similarity between the user data and each target trajectory using \textit{Rotated Correlation}. We then computed the likelihoods using the corresponding probability density function shown in Figure \ref{fig:pdfs}. Though it is similar in shape to the similarity curve, it normalises the values and pushes middling values towards the extremes of the scale. The likelihoods are then normalised into probabilities, from which we compute the overall system entropy. Initially, both the red and the green targets yield high similarities and a deterministic system would struggle to decide which to select. When the blue target reaches the edge, the uncertainty in the system further increases. This process is reflected in the entropy curve, which increases when the blue target reaches the edge, but drops when the red target reaches the side edge and its trajectory becomes substantially different to the others. Note that the likelihoods shown in the plots are instantaneous (i.e. they only consider the data in the corresponding window, ignoring previous data). The system can be further enhanced by incorporating a dynamic model that considers the previous likelihoods. This way, the probability of the blue target at point B would not be as large as the probabilities of the other targets and the overall entropy would be lower. 

\subsection{Case Study: Analysing the entropy in Orbits}

 

To understand the theoretical limits of a circular trajectory design for motion correlation, we first consider the problem of discovering the maximum number of targets we could add to our interface under the assumption that the user is able to perfectly mimic the motion of the target. Given the sampling rate of the eye tracker (30Hz), we can generate hypothetical perfect gaze trajectories with $30\times360/Speed$ samples---leaving us with trajectories of 45, 60, and 90 samples. For each of these three trajectories, we computed the similarity between all pairs of possible windows and averaged them as per the number of samples between them. This step is necessary because as we saw in Figures \ref{fig:considering_all_windows} and \ref{fig:measure_vs_angle} the correlation values vary depending on the orientation of the window relative to the coordinate system. Considering that each of these samples could be a potential target leaves us with the problem of distinguishing which of these $NSamples$ targets the user is trying to follow. As such, the initial entropy of the system is $\log_2(NSamples)$. 

Using the same approach used by Esteves et al., we thresholded the similarity values at $\lambda=0.8$. We assigned a likelihood of 1 to all windows above this threshold and zero otherwise. We then normalised the likelihoods to obtain the probabilities and the entropy. Table \ref{tab:orbits} shows our results. The table only considers targets moving in the same direction. Because in \textit{Orbits} the authors also included targets moving in the opposite direction, we can double the number of targets. We are able to make this assumption because if the target is moving in the opposite direction, at least one of the coordinate axes will be inversely correlated, ensuring that the overall correlation is negative. From the table, we see that there are on average approximately 10-16\% of windows with a similarity above the threshold depending on the overall trajectory size. In other words, there will be confusion between a target and any target within the 36-22.5 degree arc. This means that we need at least this distance between the targets, in order to minimise uncertainty. Therefore, we can calculate the maximum number of targets supported by a given design by dividing 360 by this distance, and doubling it if we also consider targets moving in the opposite direction along the same path. 

\begin{table}[ht]
\centering
\begin{tabular}{ccccc}
\toprule
 \textbf{N} & \textbf{$R>\lambda$} & \textbf{Proportion} &
 \textbf{Entropy} & \textbf{Max Targets} \\
  \midrule
45 & 7 & 0.156 & 2.81 & 2$\times$7=14 \\
60 & 9 & 0.148 & 3.17 & 2$\times$7=14 \\
90 & 9 & 0.099 & 3.17 & 2$\times$11=22 \\
\bottomrule
\end{tabular}
\caption{Simulation results for the design of \textit{Orbits}: For each trajectory size, we show the number of windows above the threshold $\lambda$, the corresponding proportion of the number of windows, the entropy of the system, and the maximum number of targets that could be placed along this trajectory in order to obtain zero entropy (we multiply this last number by 2 to consider targets moving in the opposite direction).}
\label{tab:orbits}
\end{table}

These results suggest that assuming that the user is able to follow the targets perfectly and that half of the targets will move in the opposite direction to the rest, we can add up to 14 targets with the \textit{Slow} and \textit{Medium} speeds, and up to 22 with the \textit{Fast} speed. Esteves et al.'s empirical results found acceptable error rates for up to 8 targets, but not for 16 targets, which aligns with our theoretical predictions. 

However, despite our predictions that the \textit{Fast} condition could support up to 22 targets, their study found that even 16 were too many in this condition. This is because at high speeds the eyes can no longer maintain a smooth pursuit, engaging instead in a series of saccades. This highlights the importance of user testing, as our theoretical results can only provide an upper limit for the algorithm performance, but do not say much about human capabilities without further data. Nevertheless, the advantage of a probabilistic approach is that it enables us to incorporate motor models as priors, which is a rich direction for future work.

\subsection{Effect of Noise on Entropy}
\begin{figure}[t]
    \centering
    \includegraphics[width=0.8\columnwidth]{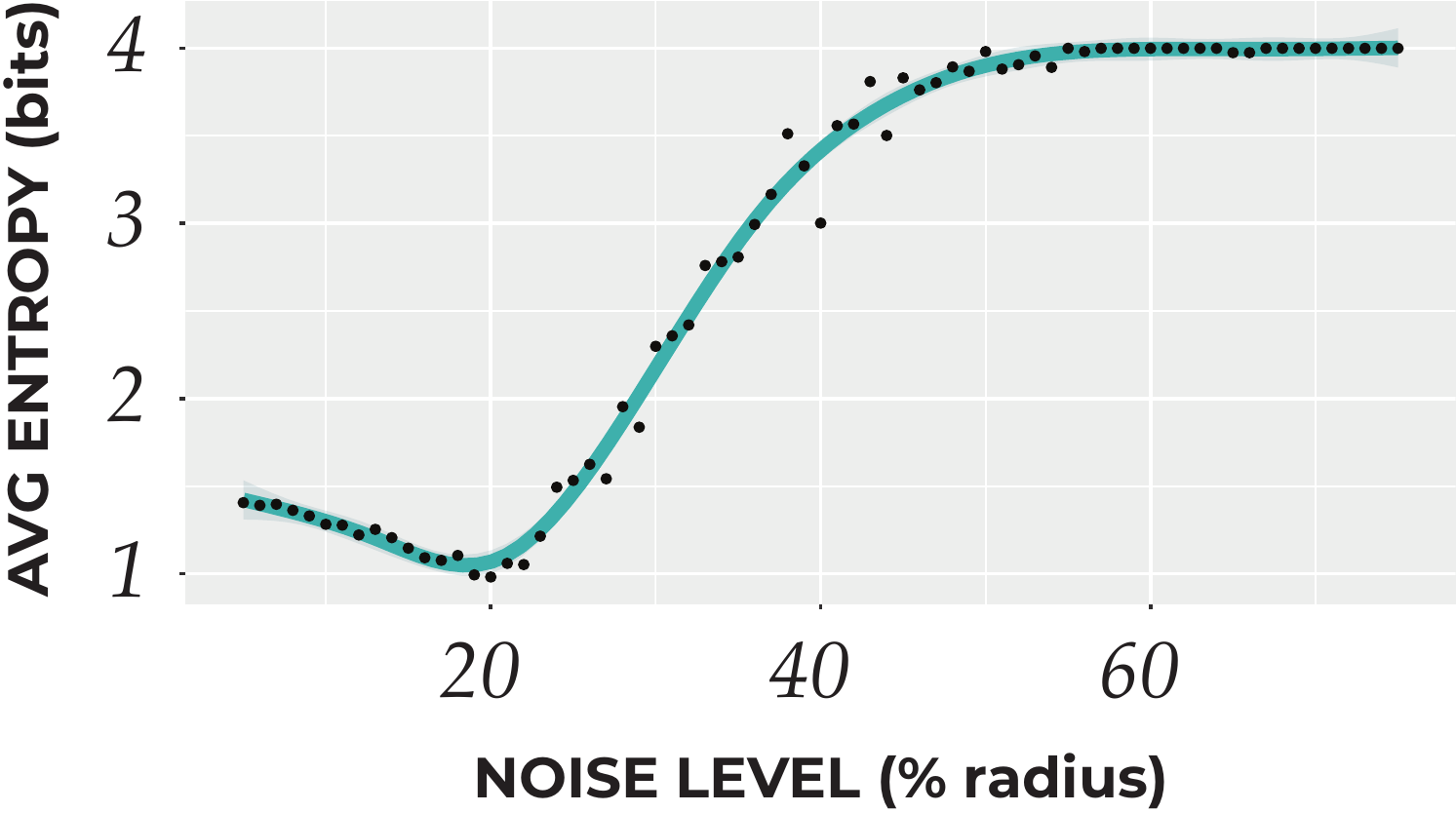}
    \caption{Effect of noise on the entropy of the system, considering 16 targets moving with the Medium speed tested in \textit{Orbits}, with a window size of 30 samples.}
    \label{fig:noise_entropy}
\end{figure}

To better understand how a design will handle noisy data, we can repeat the tests above under different noise conditions. As an example, consider Figure \ref{fig:noise_entropy}. To generate this plot, we simulated 16 targets moving around a circle at 180 degrees/s (the \textit{Medium} speed tested in \textit{Orbits}). We then simulated user input by adding Gaussian noise to the trajectory data, varying the noise between 5-75\% of the radius of the circle. We repeated each simulation 30 times and computed the average entropy across all simulations for each level of noise. The smooth curve in the figure was computed with a cubic regression spline on the data. 

The higher the noise level, the lower the correlation with the trajectory data. At small noise levels, this reduces the similarity with incorrect targets, leaving fewer targets above the threshold, and therefore reducing the uncertainty. At approximately 20\%, however, the uncertainty begins to increase, culminating at an entropy of 4, which is the maximum for 16 targets---$\log_2(16) = 4$. This is not a universal rule, and these results should be computed for each specific combination of variables, but one can use a curve like this to predict the uncertainty in the system given a sensor's noise characteristics. Conversely, given the sensor's noise characteristics, we can specify the minimum radius of the targets to achieve a desired entropy level.

\section{Discussion}



One main contribution of this paper is the re-framing of previous works on motion correlation-based interaction techniques under a probabilistic framework. Along this process, we have shown that individual design decisions in a motion correlation interface, such as which window size or similarity measure to use, should not be made in isolation as they influence each other.
Computational techniques drawn from probability and information theory are particularly useful in helping us to quantify these decisions. A probabilistic approach offers a common language that
allow us to consolidate the existing knowledge within motion correlation and also to relate it to other types of interaction techniques. In particular, it provides us with a notation to formalise the principles behind motion correlation interfaces.

Another contribution is the description of practical examples 
that demonstrates that a probabilistic approach can help us analyse motion correlation designs and understand why they work (or do not).
Next we summarise other relevant outcomes spread along this paper as a recommendation list for future work.

\subsection{Lessons Learned}

\mbox{}

\textbf{Probabilistic modelling extends previous approaches, while still being compatible with them.} Deterministic approaches are simply particular cases where the probabilities are 1 and 0. By explicitly modelling uncertainty in a system we can yield valuable design insights, such as quantifying how distinguishable each segment of the path is, understanding the limitations of different similarity measures, and enabling the integration of different sources of information.

\textbf{Interpret the similarity measure as an estimate of mutual information.} Because selection by motion correlation can be considered as transmission of information about the target trajectory, we argued that measuring motion similarities is related to estimating mutual information. As a consequence, a good similarity measure must not only yield high values when user and target motions are similar, but also yield low values in every other case. 

\textbf{Make explicit the similarity measure assumptions:} Different similarity measures perform differently depending on the assumptions behind the design. By making these assumptions explicit (for example, invariance to scale, rotation, and/or translation), we can make more informed comparisons between measures. As such, when proposing novel similarity measures, future works should clearly label which transformations are necessary to unambiguously transform the trajectory data into the gaze data.

\textbf{Use softer functions instead of step functions.} Hard thresholds make comparisons unstable when the results are near the threshold and are more vulnerable to confusion when multiple targets yield similarities above the threshold. Using softer thresholds mitigates this problem. Even better, we can incorporate our knowledge of the problem domain by using the probability density functions obtained by a mix of simulations and data collections (see Figure \ref{fig:pdfs}).

\textbf{Consider the similarity values together instead of independently.} Similarity values only consider each possible state individually, but high correlations with incorrect targets can still be consistent with perfectly following the correct target (see Figure \ref{fig:considering_all_windows}). Calculating the entropy of the probabilities of different states enables us to quantify the ambiguity of the inference and make more informed decisions about how to proceed.
 
\textbf{Understand how the entropy of the design behaves.} The shape of the trajectory, the number of samples, the window size, and the similarity measures all influence each other. Computing the inherent entropy of each window in relation to other windows helps us quantify the uncertainty of the design prior to even collecting user data and identify points of risk (see Figure \ref{fig:entropy_w}).


\textbf{Simulation is not an excuse for not collecting user data.} Though simulations can be very valuable in explicating design decisions, they do not replace user testing. However, as the community makes progress in modelling user behaviour from a computational perspective, we can expect to see more accurate models that will further empower our own simulations.

\section{Conclusion}
This paper frames interaction techniques based on motion correlation as a probabilistic reasoning problem. We argued that previous techniques can be formulated as such, and demonstrated the kinds of analysis that this formalism enables. In particular, we demonstrated analyses of similarity measures in terms of their assumptions about user and interface data, of their behaviour along different trajectory shapes, and of their value as a measure of mutual information.  We also made a case for computing conditional probabilities for each potential state and for making decisions based on entropy values rather than on independent similarities. Finally, we argued for the use of entropy as a tool for understanding the information content of a given design. 

We highlight a few practical pieces of advice for developers that stem from our analysis: (1) use the advice in section 4 to select an appropriate similarity metric, (2) collect null data to derive probability density functions for the priors (e.g. Figure \ref{fig:pdfs}), (3) select an appropriate window size and path shape based on the analysis of the entropy of the shape, (4) use entropy or other measures of uncertainty to decide whether to make a selection. All of these are ideas that can be readily incorporated in the development of systems like the ones described in the literature.
We encourage future work to frame their results in probabilistic terms in order to better contextualise their contributions and more consistently measure progress. To facilitate this, we suggest a series of lessons learned that can be taken forward in future work.

\begin{acks}
This work was partially funded by a FAPESP-University of Melbourne SPRINT Grant (Project Number: 2016/10148-3). Eduardo Velloso is the recipient of an Australian Research Council Discovery Early Career Award (Project Number: DE180100315) funded by the Australian Government, and Carlos Morimoto is the recipient of FAPESP grants no. 2016/10148-3 and 2017/50121-0.
\end{acks}

\bibliographystyle{ACM-Reference-Format}
\bibliography{sample-base}


\begin{thebibliography}{36}


\ifx \showCODEN    \undefined \def \showCODEN     #1{\unskip}     \fi
\ifx \showDOI      \undefined \def \showDOI       #1{#1}\fi
\ifx \showISBNx    \undefined \def \showISBNx     #1{\unskip}     \fi
\ifx \showISBNxiii \undefined \def \showISBNxiii  #1{\unskip}     \fi
\ifx \showISSN     \undefined \def \showISSN      #1{\unskip}     \fi
\ifx \showLCCN     \undefined \def \showLCCN      #1{\unskip}     \fi
\ifx \shownote     \undefined \def \shownote      #1{#1}          \fi
\ifx \showarticletitle \undefined \def \showarticletitle #1{#1}   \fi
\ifx \showURL      \undefined \def \showURL       {\relax}        \fi
\providecommand\bibfield[2]{#2}
\providecommand\bibinfo[2]{#2}
\providecommand\natexlab[1]{#1}
\providecommand\showeprint[2][]{arXiv:#2}

\bibitem[\protect\citeauthoryear{Biswas, Aydemir, Langdon, and Godsill}{Biswas
  et~al\mbox{.}}{2013}]%
        {Biswas:2013}
\bibfield{author}{\bibinfo{person}{Pradipta Biswas},
  \bibinfo{person}{Gokcen~Aslan Aydemir}, \bibinfo{person}{Pat Langdon}, {and}
  \bibinfo{person}{Simon Godsill}.} \bibinfo{year}{2013}\natexlab{}.
\newblock \showarticletitle{Intent recognition using neural networks and Kalman
  filters}. In \bibinfo{booktitle}{\emph{International Workshop on
  Human-Computer Interaction and Knowledge Discovery in Complex, Unstructured,
  Big Data}}. \bibinfo{publisher}{Springer}, \bibinfo{address}{Maribor,
  Slovenia}, \bibinfo{pages}{112--123}.
\newblock
\urldef\tempurl%
\url{https://doi.org/10.1007/978-3-642-39146-0_11}
\showDOI{\tempurl}


\bibitem[\protect\citeauthoryear{Carter, Velloso, Downs, Sellen, O'Hara, and
  Vetere}{Carter et~al\mbox{.}}{2016}]%
        {carter2016pathsync}
\bibfield{author}{\bibinfo{person}{Marcus Carter}, \bibinfo{person}{Eduardo
  Velloso}, \bibinfo{person}{John Downs}, \bibinfo{person}{Abigail Sellen},
  \bibinfo{person}{Kenton O'Hara}, {and} \bibinfo{person}{Frank Vetere}.}
  \bibinfo{year}{2016}\natexlab{}.
\newblock \showarticletitle{PathSync: Multi-User Gestural Interaction with
  Touchless Rhythmic Path Mimicry}. In \bibinfo{booktitle}{\emph{Proceedings of
  the 2016 CHI Conference on Human Factors in Computing Systems}} (Santa Clara,
  California, USA) \emph{(\bibinfo{series}{CHI '16})}.
  \bibinfo{publisher}{ACM}, \bibinfo{address}{New York, NY, USA},
  \bibinfo{pages}{3415--3427}.
\newblock
\showISBNx{978-1-4503-3362-7}
\urldef\tempurl%
\url{https://doi.org/10.1145/2858036.2858284}
\showDOI{\tempurl}


\bibitem[\protect\citeauthoryear{Clarke, Bellino, Esteves, Velloso, and
  Gellersen}{Clarke et~al\mbox{.}}{2016}]%
        {clarke2016tracematch}
\bibfield{author}{\bibinfo{person}{Christopher Clarke},
  \bibinfo{person}{Alessio Bellino}, \bibinfo{person}{Augusto Esteves},
  \bibinfo{person}{Eduardo Velloso}, {and} \bibinfo{person}{Hans Gellersen}.}
  \bibinfo{year}{2016}\natexlab{}.
\newblock \showarticletitle{TraceMatch: A Computer Vision Technique for User
  Input by Tracing of Animated Controls}. In
  \bibinfo{booktitle}{\emph{Proceedings of the 2016 ACM International Joint
  Conference on Pervasive and Ubiquitous Computing}} (Heidelberg, Germany)
  \emph{(\bibinfo{series}{UbiComp '16})}. \bibinfo{publisher}{ACM},
  \bibinfo{address}{New York, NY, USA}, \bibinfo{pages}{298--303}.
\newblock
\showISBNx{978-1-4503-4461-6}
\urldef\tempurl%
\url{https://doi.org/10.1145/2971648.2971714}
\showDOI{\tempurl}


\bibitem[\protect\citeauthoryear{Clarke and Gellersen}{Clarke and
  Gellersen}{2017}]%
        {clarke2017matchpoint}
\bibfield{author}{\bibinfo{person}{Christopher Clarke} {and}
  \bibinfo{person}{Hans Gellersen}.} \bibinfo{year}{2017}\natexlab{}.
\newblock \showarticletitle{MatchPoint: Spontaneous Spatial Coupling of Body
  Movement for Touchless Pointing}. In \bibinfo{booktitle}{\emph{Proceedings of
  the 30th Annual ACM Symposium on User Interface Software and Technology}}
  (Qu\&\#233;bec City, QC, Canada) \emph{(\bibinfo{series}{UIST '17})}.
  \bibinfo{publisher}{ACM}, \bibinfo{address}{New York, NY, USA},
  \bibinfo{pages}{179--192}.
\newblock
\showISBNx{978-1-4503-4981-9}
\urldef\tempurl%
\url{https://doi.org/10.1145/3126594.3126626}
\showDOI{\tempurl}


\bibitem[\protect\citeauthoryear{Cootes and Taylor}{Cootes and Taylor}{2004}]%
        {Cootes:2004}
\bibfield{author}{\bibinfo{person}{Timothy~F. Cootes} {and}
  \bibinfo{person}{Cristopher~J. Taylor}.} \bibinfo{year}{2004}\natexlab{}.
\newblock \bibinfo{title}{Statistical models of appearance for computer
  vision}.
\newblock
\newblock


\bibitem[\protect\citeauthoryear{Cox, Carter, and Velloso}{Cox
  et~al\mbox{.}}{2016}]%
        {Cox2016}
\bibfield{author}{\bibinfo{person}{Travis Cox}, \bibinfo{person}{Marcus
  Carter}, {and} \bibinfo{person}{Eduardo Velloso}.}
  \bibinfo{year}{2016}\natexlab{}.
\newblock \showarticletitle{Public DisPLAY: Social Games on Interactive Public
  Screens}. In \bibinfo{booktitle}{\emph{Proceedings of the 28th Australian
  Conference on Computer-Human Interaction}} (Launceston, Tasmania, Australia)
  \emph{(\bibinfo{series}{OzCHI '16})}. \bibinfo{publisher}{ACM},
  \bibinfo{address}{New York, NY, USA}, \bibinfo{pages}{371--380}.
\newblock
\showISBNx{978-1-4503-4618-4}
\urldef\tempurl%
\url{https://doi.org/10.1145/3010915.3010917}
\showDOI{\tempurl}


\bibitem[\protect\citeauthoryear{Drewes, Khamis, and Alt}{Drewes
  et~al\mbox{.}}{2018}]%
        {drewes2018speedsandtrajectories}
\bibfield{author}{\bibinfo{person}{Heiko Drewes}, \bibinfo{person}{Mohamed
  Khamis}, {and} \bibinfo{person}{Florian Alt}.}
  \bibinfo{year}{2018}\natexlab{}.
\newblock \showarticletitle{Smooth Pursuit Target Speeds and Trajectories}. In
  \bibinfo{booktitle}{\emph{Proceedings of the 17th International Conference on
  Mobile and Ubiquitous Multimedia}} (Cairo, Egypt) \emph{(\bibinfo{series}{MUM
  2018})}. \bibinfo{publisher}{ACM}, \bibinfo{address}{New York, NY, USA},
  \bibinfo{pages}{139--146}.
\newblock
\showISBNx{978-1-4503-6594-9}
\urldef\tempurl%
\url{https://doi.org/10.1145/3282894.3282913}
\showDOI{\tempurl}


\bibitem[\protect\citeauthoryear{Drewes, Khamis, and Alt}{Drewes
  et~al\mbox{.}}{2019}]%
        {drewes2018dialplates}
\bibfield{author}{\bibinfo{person}{Heiko Drewes}, \bibinfo{person}{Mohamed
  Khamis}, {and} \bibinfo{person}{Florian Alt}.}
  \bibinfo{year}{2019}\natexlab{}.
\newblock \showarticletitle{DialPlates: Enabling Pursuits-Based User Interfaces
  with Large Target Numbers}. In \bibinfo{booktitle}{\emph{Proceedings of the
  18th International Conference on Mobile and Ubiquitous Multimedia}} (Pisa,
  Italy) \emph{(\bibinfo{series}{MUM ’19})}. \bibinfo{publisher}{Association
  for Computing Machinery}, \bibinfo{address}{New York, NY, USA}, Article
  \bibinfo{articleno}{10}, \bibinfo{numpages}{10}~pages.
\newblock
\showISBNx{9781450376242}
\urldef\tempurl%
\url{https://doi.org/10.1145/3365610.3365626}
\showDOI{\tempurl}


\bibitem[\protect\citeauthoryear{Dryden and Mardia}{Dryden and Mardia}{1998}]%
        {Dryden:98}
\bibfield{author}{\bibinfo{person}{Ian~L. Dryden} {and}
  \bibinfo{person}{Kanti~V. Mardia}.} \bibinfo{year}{1998}\natexlab{}.
\newblock \bibinfo{booktitle}{\emph{Statistical Shape Analysis}}.
\newblock \bibinfo{publisher}{John Wiley \& Sons}, \bibinfo{address}{New York,
  NY}.
\newblock


\bibitem[\protect\citeauthoryear{Esteves, Velloso, Bulling, and
  Gellersen}{Esteves et~al\mbox{.}}{2015a}]%
        {esteves2015orbitsa}
\bibfield{author}{\bibinfo{person}{Augusto Esteves}, \bibinfo{person}{Eduardo
  Velloso}, \bibinfo{person}{Andreas Bulling}, {and} \bibinfo{person}{Hans
  Gellersen}.} \bibinfo{year}{2015}\natexlab{a}.
\newblock \showarticletitle{Orbits: Enabling Gaze Interaction in Smart Watches
  Using Moving Targets}. In \bibinfo{booktitle}{\emph{Adjunct Proceedings of
  the 2015 ACM International Joint Conference on Pervasive and Ubiquitous
  Computing and Proceedings of the 2015 ACM International Symposium on Wearable
  Computers}} (Osaka, Japan) \emph{(\bibinfo{series}{UbiComp/ISWC'15
  Adjunct})}. \bibinfo{publisher}{ACM}, \bibinfo{address}{New York, NY, USA},
  \bibinfo{pages}{419--422}.
\newblock
\showISBNx{978-1-4503-3575-1}
\urldef\tempurl%
\url{https://doi.org/10.1145/2800835.2800942}
\showDOI{\tempurl}


\bibitem[\protect\citeauthoryear{Esteves, Velloso, Bulling, and
  Gellersen}{Esteves et~al\mbox{.}}{2015b}]%
        {esteves2015orbitsb}
\bibfield{author}{\bibinfo{person}{Augusto Esteves}, \bibinfo{person}{Eduardo
  Velloso}, \bibinfo{person}{Andreas Bulling}, {and} \bibinfo{person}{Hans
  Gellersen}.} \bibinfo{year}{2015}\natexlab{b}.
\newblock \showarticletitle{Orbits: Gaze Interaction for Smart Watches Using
  Smooth Pursuit Eye Movements}. In \bibinfo{booktitle}{\emph{Proceedings of
  the 28th Annual ACM Symposium on User Interface Software \& Technology}}
  (Daegu, Kyungpook, Republic of Korea) \emph{(\bibinfo{series}{UIST '15})}.
  \bibinfo{publisher}{ACM}, \bibinfo{address}{New York, NY, USA},
  \bibinfo{pages}{457--466}.
\newblock
\showISBNx{978-1-4503-3779-3}
\urldef\tempurl%
\url{https://doi.org/10.1145/2807442.2807499}
\showDOI{\tempurl}


\bibitem[\protect\citeauthoryear{Esteves, Verweij, Suraiya, Islam, Lee, and
  Oakley}{Esteves et~al\mbox{.}}{2017}]%
        {esteves2017smoothmoves}
\bibfield{author}{\bibinfo{person}{Augusto Esteves}, \bibinfo{person}{David
  Verweij}, \bibinfo{person}{Liza Suraiya}, \bibinfo{person}{Rasel Islam},
  \bibinfo{person}{Youryang Lee}, {and} \bibinfo{person}{Ian Oakley}.}
  \bibinfo{year}{2017}\natexlab{}.
\newblock \showarticletitle{SmoothMoves: Smooth Pursuits Head Movements for
  Augmented Reality}. In \bibinfo{booktitle}{\emph{Proceedings of the 30th
  Annual ACM Symposium on User Interface Software and Technology}}
  (Qu\&\#233;bec City, QC, Canada) \emph{(\bibinfo{series}{UIST '17})}.
  \bibinfo{publisher}{ACM}, \bibinfo{address}{New York, NY, USA},
  \bibinfo{pages}{167--178}.
\newblock
\showISBNx{978-1-4503-4981-9}
\urldef\tempurl%
\url{https://doi.org/10.1145/3126594.3126616}
\showDOI{\tempurl}


\bibitem[\protect\citeauthoryear{Fekete, Elmqvist, and Guiard}{Fekete
  et~al\mbox{.}}{2009}]%
        {fekete2009motion}
\bibfield{author}{\bibinfo{person}{Jean-Daniel Fekete}, \bibinfo{person}{Niklas
  Elmqvist}, {and} \bibinfo{person}{Yves Guiard}.}
  \bibinfo{year}{2009}\natexlab{}.
\newblock \showarticletitle{Motion-pointing: Target Selection Using Elliptical
  Motions}. In \bibinfo{booktitle}{\emph{Proceedings of the SIGCHI Conference
  on Human Factors in Computing Systems}} (Boston, MA, USA)
  \emph{(\bibinfo{series}{CHI '09})}. \bibinfo{publisher}{ACM},
  \bibinfo{address}{New York, NY, USA}, \bibinfo{pages}{289--298}.
\newblock
\showISBNx{978-1-60558-246-7}
\urldef\tempurl%
\url{https://doi.org/10.1145/1518701.1518748}
\showDOI{\tempurl}


\bibitem[\protect\citeauthoryear{Gomez and Gellersen}{Gomez and
  Gellersen}{2018}]%
        {gomez2018smoothi}
\bibfield{author}{\bibinfo{person}{Argenis~Ramirez Gomez} {and}
  \bibinfo{person}{Hans Gellersen}.} \bibinfo{year}{2018}\natexlab{}.
\newblock \showarticletitle{Smooth-i: Smart Re-calibration Using Smooth Pursuit
  Eye Movements}. In \bibinfo{booktitle}{\emph{Proceedings of the 2018 ACM
  Symposium on Eye Tracking Research \& Applications}} (Warsaw, Poland)
  \emph{(\bibinfo{series}{ETRA '18})}. \bibinfo{publisher}{ACM},
  \bibinfo{address}{New York, NY, USA}, Article \bibinfo{articleno}{10},
  \bibinfo{numpages}{5}~pages.
\newblock
\showISBNx{978-1-4503-5706-7}
\urldef\tempurl%
\url{https://doi.org/10.1145/3204493.3204585}
\showDOI{\tempurl}


\bibitem[\protect\citeauthoryear{Goodwin and Leech}{Goodwin and Leech}{2006}]%
        {goodwin2006correlation}
\bibfield{author}{\bibinfo{person}{Laura~D. Goodwin} {and}
  \bibinfo{person}{Nancy~L. Leech}.} \bibinfo{year}{2006}\natexlab{}.
\newblock \showarticletitle{Understanding Correlation: Factors That Affect the
  Size of r}.
\newblock \bibinfo{journal}{\emph{The Journal of Experimental Education}}
  \bibinfo{volume}{74}, \bibinfo{number}{3} (\bibinfo{year}{2006}),
  \bibinfo{pages}{251--266}.
\newblock
\showISSN{00220973, 19400683}
\urldef\tempurl%
\url{https://doi.org/10.3200/JEXE.74.3.249-266}
\showDOI{\tempurl}


\bibitem[\protect\citeauthoryear{Hornb{\ae}k and Oulasvirta}{Hornb{\ae}k and
  Oulasvirta}{2017}]%
        {hornbaek2017whatisinteraction}
\bibfield{author}{\bibinfo{person}{Kasper Hornb{\ae}k} {and}
  \bibinfo{person}{Antti Oulasvirta}.} \bibinfo{year}{2017}\natexlab{}.
\newblock \showarticletitle{What Is Interaction?}. In
  \bibinfo{booktitle}{\emph{Proceedings of the 2017 CHI Conference on Human
  Factors in Computing Systems}} (Denver, Colorado, USA)
  \emph{(\bibinfo{series}{CHI '17})}. \bibinfo{publisher}{ACM},
  \bibinfo{address}{New York, NY, USA}, \bibinfo{pages}{5040--5052}.
\newblock
\showISBNx{978-1-4503-4655-9}
\urldef\tempurl%
\url{https://doi.org/10.1145/3025453.3025765}
\showDOI{\tempurl}


\bibitem[\protect\citeauthoryear{Khamis, Oechsner, Alt, and Bulling}{Khamis
  et~al\mbox{.}}{2018}]%
        {khamis2018vrpursuits}
\bibfield{author}{\bibinfo{person}{Mohamed Khamis}, \bibinfo{person}{Carl
  Oechsner}, \bibinfo{person}{Florian Alt}, {and} \bibinfo{person}{Andreas
  Bulling}.} \bibinfo{year}{2018}\natexlab{}.
\newblock \showarticletitle{VRpursuits: Interaction in Virtual Reality Using
  Smooth Pursuit Eye Movements}. In \bibinfo{booktitle}{\emph{Proceedings of
  the 2018 International Conference on Advanced Visual Interfaces}}
  (Castiglione della Pescaia, Grosseto, Italy) \emph{(\bibinfo{series}{AVI
  '18})}. \bibinfo{publisher}{ACM}, \bibinfo{address}{New York, NY, USA},
  Article \bibinfo{articleno}{18}, \bibinfo{numpages}{8}~pages.
\newblock
\showISBNx{978-1-4503-5616-9}
\urldef\tempurl%
\url{https://doi.org/10.1145/3206505.3206522}
\showDOI{\tempurl}


\bibitem[\protect\citeauthoryear{Khamis, Saltuk, Hang, Stolz, Bulling, and
  Alt}{Khamis et~al\mbox{.}}{2016}]%
        {khamis2016textpursuits}
\bibfield{author}{\bibinfo{person}{Mohamed Khamis}, \bibinfo{person}{Ozan
  Saltuk}, \bibinfo{person}{Alina Hang}, \bibinfo{person}{Katharina Stolz},
  \bibinfo{person}{Andreas Bulling}, {and} \bibinfo{person}{Florian Alt}.}
  \bibinfo{year}{2016}\natexlab{}.
\newblock \showarticletitle{TextPursuits: Using Text for Pursuits-based
  Interaction and Calibration on Public Displays}. In
  \bibinfo{booktitle}{\emph{Proceedings of the 2016 ACM International Joint
  Conference on Pervasive and Ubiquitous Computing}} (Heidelberg, Germany)
  \emph{(\bibinfo{series}{UbiComp '16})}. \bibinfo{publisher}{ACM},
  \bibinfo{address}{New York, NY, USA}, \bibinfo{pages}{274--285}.
\newblock
\showISBNx{978-1-4503-4461-6}
\urldef\tempurl%
\url{https://doi.org/10.1145/2971648.2971679}
\showDOI{\tempurl}


\bibitem[\protect\citeauthoryear{Langford, Schwertman, and Owens}{Langford
  et~al\mbox{.}}{2001}]%
        {langford2001correlationtransitivity}
\bibfield{author}{\bibinfo{person}{Eric Langford}, \bibinfo{person}{Neil
  Schwertman}, {and} \bibinfo{person}{Margaret Owens}.}
  \bibinfo{year}{2001}\natexlab{}.
\newblock \showarticletitle{Is the Property of Being Positively Correlated
  Transitive?}
\newblock \bibinfo{journal}{\emph{The American Statistician}}
  \bibinfo{volume}{55}, \bibinfo{number}{4} (\bibinfo{year}{2001}),
  \bibinfo{pages}{322--325}.
\newblock
\urldef\tempurl%
\url{https://doi.org/10.1198/000313001753272286}
\showDOI{\tempurl}
\showeprint{https://doi.org/10.1198/000313001753272286}


\bibitem[\protect\citeauthoryear{Liu, D'Oliveira, Beaudouin-Lafon, and
  Rioul}{Liu et~al\mbox{.}}{2017}]%
        {liu2017bignav}
\bibfield{author}{\bibinfo{person}{Wanyu Liu}, \bibinfo{person}{Rafael~Lucas
  D'Oliveira}, \bibinfo{person}{Michel Beaudouin-Lafon}, {and}
  \bibinfo{person}{Olivier Rioul}.} \bibinfo{year}{2017}\natexlab{}.
\newblock \showarticletitle{BIGnav: Bayesian Information Gain for Guiding
  Multiscale Navigation}. In \bibinfo{booktitle}{\emph{Proceedings of the 2017
  CHI Conference on Human Factors in Computing Systems}} (Denver, Colorado,
  USA) \emph{(\bibinfo{series}{CHI '17})}. \bibinfo{publisher}{Association for
  Computing Machinery}, \bibinfo{address}{New York, NY, USA},
  \bibinfo{pages}{5869–5880}.
\newblock
\showISBNx{9781450346559}
\urldef\tempurl%
\url{https://doi.org/10.1145/3025453.3025524}
\showDOI{\tempurl}


\bibitem[\protect\citeauthoryear{Liu, Rioul, McGrenere, Mackay, and
  Beaudouin-Lafon}{Liu et~al\mbox{.}}{2018}]%
        {liu2018bigfile}
\bibfield{author}{\bibinfo{person}{Wanyu Liu}, \bibinfo{person}{Olivier Rioul},
  \bibinfo{person}{Joanna McGrenere}, \bibinfo{person}{Wendy~E. Mackay}, {and}
  \bibinfo{person}{Michel Beaudouin-Lafon}.} \bibinfo{year}{2018}\natexlab{}.
\newblock \showarticletitle{BIGFile: Bayesian Information Gain for Fast File
  Retrieval}. In \bibinfo{booktitle}{\emph{Proceedings of the 2018 CHI
  Conference on Human Factors in Computing Systems}} (Montreal QC, Canada)
  \emph{(\bibinfo{series}{CHI '18})}. \bibinfo{publisher}{Association for
  Computing Machinery}, \bibinfo{address}{New York, NY, USA},
  \bibinfo{pages}{1–13}.
\newblock
\showISBNx{9781450356206}
\urldef\tempurl%
\url{https://doi.org/10.1145/3173574.3173959}
\showDOI{\tempurl}


\bibitem[\protect\citeauthoryear{MacKay}{MacKay}{2003}]%
        {mackay2003information}
\bibfield{author}{\bibinfo{person}{David~J.C. MacKay}.}
  \bibinfo{year}{2003}\natexlab{}.
\newblock \bibinfo{booktitle}{\emph{Information theory, inference and learning
  algorithms}}.
\newblock \bibinfo{publisher}{Cambridge university press},
  \bibinfo{address}{United Kingdom}.
\newblock
\urldef\tempurl%
\url{https://doi.org/10.2277/0521642981}
\showDOI{\tempurl}


\bibitem[\protect\citeauthoryear{Oulasvirta, Bi, and Howes}{Oulasvirta
  et~al\mbox{.}}{2018}]%
        {oulasvirta2018computational}
\bibfield{author}{\bibinfo{person}{Antti Oulasvirta}, \bibinfo{person}{Xiaojun
  Bi}, {and} \bibinfo{person}{Andrew Howes}.} \bibinfo{year}{2018}\natexlab{}.
\newblock \bibinfo{booktitle}{\emph{Computational interaction}}.
\newblock \bibinfo{publisher}{Oxford University Press},
  \bibinfo{address}{United Kingdom}.
\newblock
\urldef\tempurl%
\url{https://doi.org/10.1093/oso/9780198799603.001.0001}
\showDOI{\tempurl}


\bibitem[\protect\citeauthoryear{Pfeuffer, Vidal, Turner, Bulling, and
  Gellersen}{Pfeuffer et~al\mbox{.}}{2013}]%
        {pfeuffer2013pursuit}
\bibfield{author}{\bibinfo{person}{Ken Pfeuffer}, \bibinfo{person}{M{\'e}lodie
  Vidal}, \bibinfo{person}{Jayson Turner}, \bibinfo{person}{Andreas Bulling},
  {and} \bibinfo{person}{Hans Gellersen}.} \bibinfo{year}{2013}\natexlab{}.
\newblock \showarticletitle{Pursuit Calibration: Making Gaze Calibration Less
  Tedious and More Flexible}. In \bibinfo{booktitle}{\emph{Proceedings of the
  26th Annual ACM Symposium on User Interface Software and Technology}} (St.
  Andrews, Scotland, United Kingdom) \emph{(\bibinfo{series}{UIST '13})}.
  \bibinfo{publisher}{ACM}, \bibinfo{address}{New York, NY, USA},
  \bibinfo{pages}{261--270}.
\newblock
\showISBNx{978-1-4503-2268-3}
\urldef\tempurl%
\url{https://doi.org/10.1145/2501988.2501998}
\showDOI{\tempurl}


\bibitem[\protect\citeauthoryear{Rogers, Williamson, Stewart, and
  Murray-Smith}{Rogers et~al\mbox{.}}{2010}]%
        {rogers2010fingercloud}
\bibfield{author}{\bibinfo{person}{Simon Rogers}, \bibinfo{person}{John
  Williamson}, \bibinfo{person}{Craig Stewart}, {and} \bibinfo{person}{Roderick
  Murray-Smith}.} \bibinfo{year}{2010}\natexlab{}.
\newblock \showarticletitle{FingerCloud: Uncertainty and Autonomy Handover
  Incapacitive Sensing}. In \bibinfo{booktitle}{\emph{Proceedings of the SIGCHI
  Conference on Human Factors in Computing Systems}} (Atlanta, Georgia, USA)
  \emph{(\bibinfo{series}{CHI '10})}. \bibinfo{publisher}{Association for
  Computing Machinery}, \bibinfo{address}{New York, NY, USA},
  \bibinfo{pages}{577–580}.
\newblock
\showISBNx{9781605589299}
\urldef\tempurl%
\url{https://doi.org/10.1145/1753326.1753412}
\showDOI{\tempurl}


\bibitem[\protect\citeauthoryear{Schwarz, Hudson, Mankoff, and Wilson}{Schwarz
  et~al\mbox{.}}{2010}]%
        {schwarz2010framework}
\bibfield{author}{\bibinfo{person}{Julia Schwarz}, \bibinfo{person}{Scott
  Hudson}, \bibinfo{person}{Jennifer Mankoff}, {and} \bibinfo{person}{Andrew~D.
  Wilson}.} \bibinfo{year}{2010}\natexlab{}.
\newblock \showarticletitle{A Framework for Robust and Flexible Handling of
  Inputs with Uncertainty}. In \bibinfo{booktitle}{\emph{Proceedings of the
  23nd Annual ACM Symposium on User Interface Software and Technology}}
  \emph{(\bibinfo{series}{UIST ’10})}. \bibinfo{publisher}{Association for
  Computing Machinery}, \bibinfo{address}{New York, NY},
  \bibinfo{pages}{47–56}.
\newblock
\showISBNx{9781450302715}
\urldef\tempurl%
\url{https://doi.org/10.1145/1866029.1866039}
\showDOI{\tempurl}


\bibitem[\protect\citeauthoryear{Shi, Gunn, and Damper}{Shi
  et~al\mbox{.}}{2003}]%
        {shi2003handwritten}
\bibfield{author}{\bibinfo{person}{Daming Shi}, \bibinfo{person}{Steve~R.
  Gunn}, {and} \bibinfo{person}{Robert~I. Damper}.}
  \bibinfo{year}{2003}\natexlab{}.
\newblock \showarticletitle{Handwritten Chinese radical recognition using
  nonlinear active shape models}.
\newblock \bibinfo{journal}{\emph{IEEE transactions on pattern analysis and
  machine intelligence}} \bibinfo{volume}{25}, \bibinfo{number}{2}
  (\bibinfo{year}{2003}), \bibinfo{pages}{277--280}.
\newblock
\urldef\tempurl%
\url{https://doi.org/10.1109/TPAMI.2003.1177158}
\showDOI{\tempurl}


\bibitem[\protect\citeauthoryear{Velloso, Carter, Newn, Esteves, Clarke, and
  Gellersen}{Velloso et~al\mbox{.}}{2017}]%
        {velloso2017motioncorrelation}
\bibfield{author}{\bibinfo{person}{Eduardo Velloso}, \bibinfo{person}{Marcus
  Carter}, \bibinfo{person}{Joshua Newn}, \bibinfo{person}{Augusto Esteves},
  \bibinfo{person}{Christopher Clarke}, {and} \bibinfo{person}{Hans
  Gellersen}.} \bibinfo{year}{2017}\natexlab{}.
\newblock \showarticletitle{Motion Correlation: Selecting Objects by Matching
  Their Movement}.
\newblock \bibinfo{journal}{\emph{ACM Trans. Comput.-Hum. Interact.}}
  \bibinfo{volume}{24}, \bibinfo{number}{3}, Article \bibinfo{articleno}{22}
  (\bibinfo{date}{April} \bibinfo{year}{2017}), \bibinfo{numpages}{35}~pages.
\newblock
\showISSN{1073-0516}
\urldef\tempurl%
\url{https://doi.org/10.1145/3064937}
\showDOI{\tempurl}


\bibitem[\protect\citeauthoryear{Velloso, Coutinho, Kurauchi, and
  Morimoto}{Velloso et~al\mbox{.}}{2018}]%
        {velloso2018circular}
\bibfield{author}{\bibinfo{person}{Eduardo Velloso},
  \bibinfo{person}{Flavio~Luiz Coutinho}, \bibinfo{person}{Andrew Kurauchi},
  {and} \bibinfo{person}{Carlos~H Morimoto}.} \bibinfo{year}{2018}\natexlab{}.
\newblock \showarticletitle{Circular Orbits Detection for Gaze Interaction
  Using 2D Correlation and Profile Matching Algorithms}. In
  \bibinfo{booktitle}{\emph{Proceedings of the 2018 ACM Symposium on Eye
  Tracking Research \& Applications}} (Warsaw, Poland)
  \emph{(\bibinfo{series}{ETRA '18})}. \bibinfo{publisher}{ACM},
  \bibinfo{address}{New York, NY, USA}, Article \bibinfo{articleno}{25},
  \bibinfo{numpages}{9}~pages.
\newblock
\showISBNx{978-1-4503-5706-7}
\urldef\tempurl%
\url{https://doi.org/10.1145/3204493.3204524}
\showDOI{\tempurl}


\bibitem[\protect\citeauthoryear{Velloso, Wirth, Weichel, Esteves, and
  Gellersen}{Velloso et~al\mbox{.}}{2016}]%
        {velloso2016ambigaze}
\bibfield{author}{\bibinfo{person}{Eduardo Velloso}, \bibinfo{person}{Markus
  Wirth}, \bibinfo{person}{Christian Weichel}, \bibinfo{person}{Augusto
  Esteves}, {and} \bibinfo{person}{Hans Gellersen}.}
  \bibinfo{year}{2016}\natexlab{}.
\newblock \showarticletitle{AmbiGaze: Direct Control of Ambient Devices by
  Gaze}. In \bibinfo{booktitle}{\emph{Proceedings of the 2016 ACM Conference on
  Designing Interactive Systems}} (Brisbane, QLD, Australia)
  \emph{(\bibinfo{series}{DIS '16})}. \bibinfo{publisher}{ACM},
  \bibinfo{address}{New York, NY, USA}, \bibinfo{pages}{812--817}.
\newblock
\showISBNx{978-1-4503-4031-1}
\urldef\tempurl%
\url{https://doi.org/10.1145/2901790.2901867}
\showDOI{\tempurl}


\bibitem[\protect\citeauthoryear{Verweij, Esteves, Khan, and Bakker}{Verweij
  et~al\mbox{.}}{2017}]%
        {verweij2017wavetrace}
\bibfield{author}{\bibinfo{person}{David Verweij}, \bibinfo{person}{Augusto
  Esteves}, \bibinfo{person}{Vassilis-Javed Khan}, {and}
  \bibinfo{person}{Saskia Bakker}.} \bibinfo{year}{2017}\natexlab{}.
\newblock \showarticletitle{WaveTrace: Motion Matching Input Using Wrist-Worn
  Motion Sensors}. In \bibinfo{booktitle}{\emph{Proceedings of the 2017 CHI
  Conference Extended Abstracts on Human Factors in Computing Systems}}
  (Denver, Colorado, USA) \emph{(\bibinfo{series}{CHI EA '17})}.
  \bibinfo{publisher}{ACM}, \bibinfo{address}{New York, NY, USA},
  \bibinfo{pages}{2180--2186}.
\newblock
\showISBNx{978-1-4503-4656-6}
\urldef\tempurl%
\url{https://doi.org/10.1145/3027063.3053161}
\showDOI{\tempurl}


\bibitem[\protect\citeauthoryear{Vidal, Bulling, and Gellersen}{Vidal
  et~al\mbox{.}}{2013}]%
        {vidal2013pursuitsb}
\bibfield{author}{\bibinfo{person}{M{\'e}lodie Vidal}, \bibinfo{person}{Andreas
  Bulling}, {and} \bibinfo{person}{Hans Gellersen}.}
  \bibinfo{year}{2013}\natexlab{}.
\newblock \showarticletitle{Pursuits: Spontaneous Interaction with Displays
  Based on Smooth Pursuit Eye Movement and Moving Targets}. In
  \bibinfo{booktitle}{\emph{Proceedings of the 2013 ACM International Joint
  Conference on Pervasive and Ubiquitous Computing}} (Zurich, Switzerland)
  \emph{(\bibinfo{series}{UbiComp '13})}. \bibinfo{publisher}{ACM},
  \bibinfo{address}{New York, NY, USA}, \bibinfo{pages}{439--448}.
\newblock
\showISBNx{978-1-4503-1770-2}
\urldef\tempurl%
\url{https://doi.org/10.1145/2493432.2493477}
\showDOI{\tempurl}


\bibitem[\protect\citeauthoryear{Walters-Williams and Li}{Walters-Williams and
  Li}{2009}]%
        {walters2009mutualinformation}
\bibfield{author}{\bibinfo{person}{Janett Walters-Williams} {and}
  \bibinfo{person}{Yan Li}.} \bibinfo{year}{2009}\natexlab{}.
\newblock \showarticletitle{Estimation of Mutual Information: A Survey}. In
  \bibinfo{booktitle}{\emph{Rough Sets and Knowledge Technology}},
  \bibfield{editor}{\bibinfo{person}{Peng Wen}, \bibinfo{person}{Yuefeng Li},
  \bibinfo{person}{Lech Polkowski}, \bibinfo{person}{Yiyu Yao},
  \bibinfo{person}{Shusaku Tsumoto}, {and} \bibinfo{person}{Guoyin Wang}}
  (Eds.). \bibinfo{publisher}{Springer Berlin Heidelberg},
  \bibinfo{address}{Berlin, Heidelberg}, \bibinfo{pages}{389--396}.
\newblock
\showISBNx{978-3-642-02962-2}
\urldef\tempurl%
\url{https://doi.org/10.1007/978-3-642-02962-2_49}
\showDOI{\tempurl}


\bibitem[\protect\citeauthoryear{Williamson}{Williamson}{2006}]%
        {williamson2006continuous}
\bibfield{author}{\bibinfo{person}{John Williamson}.}
  \bibinfo{year}{2006}\natexlab{}.
\newblock \emph{\bibinfo{title}{Continuous uncertain interaction}}.
\newblock \bibinfo{thesistype}{Ph.D. Dissertation}. \bibinfo{school}{University
  of Glasgow}.
\newblock


\bibitem[\protect\citeauthoryear{Williamson and Murray-Smith}{Williamson and
  Murray-Smith}{2004}]%
        {williamson2004pointing}
\bibfield{author}{\bibinfo{person}{John Williamson} {and}
  \bibinfo{person}{Roderick Murray-Smith}.} \bibinfo{year}{2004}\natexlab{}.
\newblock \showarticletitle{Pointing Without a Pointer}. In
  \bibinfo{booktitle}{\emph{CHI '04 Extended Abstracts on Human Factors in
  Computing Systems}} (Vienna, Austria) \emph{(\bibinfo{series}{CHI EA '04})}.
  \bibinfo{publisher}{ACM}, \bibinfo{address}{New York, NY, USA},
  \bibinfo{pages}{1407--1410}.
\newblock
\showISBNx{1-58113-703-6}
\urldef\tempurl%
\url{https://doi.org/10.1145/985921.986076}
\showDOI{\tempurl}


\bibitem[\protect\citeauthoryear{Zhang, Wang, Waghmare, Jain, Ploetz, Inan,
  Starner, and Abowd}{Zhang et~al\mbox{.}}{2017}]%
        {zhang2017fingorbits}
\bibfield{author}{\bibinfo{person}{Cheng Zhang}, \bibinfo{person}{Xiaoxuan
  Wang}, \bibinfo{person}{Anandghan Waghmare}, \bibinfo{person}{Sumeet Jain},
  \bibinfo{person}{Thomas Ploetz}, \bibinfo{person}{Omer~T. Inan},
  \bibinfo{person}{Thad~E. Starner}, {and} \bibinfo{person}{Gregory~D. Abowd}.}
  \bibinfo{year}{2017}\natexlab{}.
\newblock \showarticletitle{FingOrbits: Interaction with Wearables Using
  Synchronized Thumb Movements}. In \bibinfo{booktitle}{\emph{Proceedings of
  the 2017 ACM International Symposium on Wearable Computers}} (Maui, Hawaii)
  \emph{(\bibinfo{series}{ISWC '17})}. \bibinfo{publisher}{ACM},
  \bibinfo{address}{New York, NY, USA}, \bibinfo{pages}{62--65}.
\newblock
\showISBNx{978-1-4503-5188-1}
\urldef\tempurl%
\url{https://doi.org/10.1145/3123021.3123041}
\showDOI{\tempurl}


\end{thebibliography}

\appendix

\end{document}